\documentclass[final,leqno,showlabe]{siamltex}

\usepackage{stmaryrd}
\usepackage{amsmath}
\usepackage{amssymb}
\usepackage{cite}
\usepackage{caption}
\usepackage{tabularx}
\usepackage{color}

\input psfig.sty

\newcommand{\p}{\partial}
\newcommand{\og}{\omega}
\newcommand{\Og}{\Omega}
\newcommand{\fl}[2]{\frac{#1}{#2}}
\newcommand{\dt}{\delta}
\newcommand{\nn}{\nonumber}
\newcommand{\ap}{\alpha}
\newcommand{\bt}{\beta}

\newcommand{\Dt}{\Delta}
\renewcommand{\theequation}{\arabic{section}.\arabic{equation}}
\newcommand{\be}{\begin{equation}}
\newcommand{\ee}{\end{equation}}
\newcommand{\ba}{\begin{array}}
\newcommand{\ea}{\end{array}}
\newcommand{\bea}{\begin{eqnarray}}
\newcommand{\eea}{\end{eqnarray}}
\newcommand{\beas}{\begin{eqnarray*}}
\newcommand{\eeas}{\end{eqnarray*}}
\newtheorem{remark}{Remark}[section]
\newcommand{\bx}{{\bf x} }
\newcommand{\gm}{\gamma}
\newcommand{\vphi}{\varphi}
\newcommand{\tbx}{\widetilde{\bf x} }
\newcommand{\bn}{{\bf n} }
\newcommand{\tx}{{\widetilde x} }
\newcommand{\ty}{{\widetilde y} }
\newcommand{\tz}{{\widetilde z} }

\newcommand{\R}{{\Bbb R}}

\title{A simple and efficient numerical method for computing the dynamics of rotating
 Bose-Einstein condensates via a rotating Lagrangian coordinate\thanks{This
research was supported by the Singapore A*STAR SERC ``Complex Systems'' Research Programme
grant 1224504056 (W. Bao and Q. Tang), a research award by the
King Abdullah University of Science and Technology (KAUST) No. KUK-I1-007-43 (D. Marahrens), and
the University of Missouri Research Board and the Simons Foundation Award No. 210138 (Y. Zhang).}}

\author{Weizhu Bao\thanks{Department of Mathematics and Center for Computational Science and
Engineering,  National University of Singapore, Singapore 119076 (URL: {http://www.math.nus.edu.sg/\~{}bao/}, {\tt bao@math.nus.edu.sg} (W. Bao); {\tt g0800880@nus.edu.sg} (Q. Tang)).}\and
Daniel Marahrens\thanks{Max Planck Institute for Mathematics in the Sciences, Inselstr.\ 22, 04103 Leipzig, Germany
 ({\tt Daniel.Marahrens@mis.mpg.de}).} \and
Qinglin Tang\footnotemark[2] \and
Yanzhi Zhang\thanks{Department of Mathematics and Statistics, Missouri University of Science
and Technology, Rolla, MO
65409-0020, USA ({\tt zhangyanz@mst.edu}).}}
\date{}

\begin{document}

\maketitle

\begin{abstract}
We propose a simple, efficient and accurate numerical method for simulating the dynamics of rotating
 Bose-Einstein condensates (BECs) in a rotational frame with/without a long-range
dipole-dipole interaction. We begin with the three-dimensional (3D) Gross-Pitaevskii equation (GPE)
with an angular momentum rotation term and/or long-range dipole-dipole interaction,
state the two-dimensional (2D) GPE obtained from the 3D GPE via dimension reduction under anisotropic
external potential and review some dynamical laws related to the 2D and 3D GPE.
By introducing a rotating Lagrangian coordinate system, the original GPEs are re-formulated to GPEs
without the angular momentum rotation which is replaced by a time-dependent potential in the new
coordinate system. We then cast the conserved quantities and dynamical laws in
the new rotating Lagrangian coordinates. Based on the new formulation of the GPE for rotating BECs
 in the rotating Lagrangian coordinates, a time-splitting spectral method is presented for computing the dynamics of rotating BECs. The new numerical method is explicit, simple to implement, unconditionally stable and very efficient in computation. It is spectral order accurate in space and second-order
accurate in time, and conserves the mass in the discrete level. Extensive numerical results
are reported to demonstrate the efficiency and accuracy of the new numerical method. Finally,
the numerical method is applied to test the dynamical laws of rotating BECs such as the dynamics of
condensate width, angular momentum expectation and center-of-mass, and to investigate numerically
the dynamics and interaction of quantized vortex lattices
in rotating BECs without/with the long-range dipole-dipole interaction.
\end{abstract}

\begin{keywords} Rotating Bose-Einstein condensate, dipole-dipole interaction,
Gross-Pitaevskii equation, angular momentum rotation, rotating
Lagrangian coordinates, time-splitting.
\end{keywords}

\begin{AMS}
35Q41, 65M70, 81Q05, 81V45, 82D50
\end{AMS}

\pagestyle{myheadings}\thispagestyle{plain}
 \markboth{W. Bao, D. Marahrens, Q. Tang and Y. Zhang}
{METHOD FOR ROTATING BOSE-EINSTEIN CONDENSATES}

\section{Introduction}
\setcounter{equation}{0}
\label{section1}

Bose--Einstein condensation (BEC), first observed in 1995 \cite{Anderson,Bradley,Davis}, has provided a platform
to study the macroscopic quantum world.  Later, with the observation of quantized vortices
 \cite{Matthews1999, Madison2000, Madison2001, Raman2001,Ketterle2001,Yuce2010,Bretin2004},
 rotating BECs have been extensively studied  in the laboratory. The occurrence of
quantized vortices is a hallmark of the superfluid nature of  Bose--Einstein condensates.
In addition, condensation of bosonic atoms and molecules with significant  dipole moments
whose interaction is both nonlocal and anisotropic has recently been achieved experimentally in trapped
$^{52}$Cr and $^{164}$Dy gases \cite{Griesmaier2005, Lu2011, Lahaye2009,Abad2009,Cooper2005,Malet2011,Xiong2009}.

At temperatures $T$ much smaller than the critical  temperature $T_c$, the properties of a BEC
in a rotating frame with long-range dipole-dipole interaction are well described by the macroscopic
 complex-valued wave function $\psi=\psi(\bx,t)$, whose evolution
is governed by the three-dimensional (3D) Gross-Pitaevskii equation (GPE) in dimensionless units with angular
momentum rotation term and long-range dipole-dipole interaction
\cite{Zhang2005, Yi2006, Bijnen2009, Bao2010, Cai2010,Abad2009,Seiringer2003}:
\be\label{NGPE}
 i\p_t \psi({\bx}, t) = \left[-\fl{1}{2}\nabla^2 + V({\bf x}) + \kappa |\psi|^2 +
\lambda\left(U_{\rm dip}*|\psi|^2\right) - \Omega L_{z}\right]\psi(\bx,t),
\ee
where $t$ denotes time, ${\bf x} = (x, y, z)^T \in {\mathbb R}^3$ is the
Cartesian coordinate vector,  $V(\bx)$ is a given real-valued external trapping potential
which is determined by the type of system under investigation and
$\kappa=\frac{4\pi N a_s}{x_s}$ is a dimensionless constant  describing the strength of the short-range two-body
interaction (positive for repulsive interaction, and resp. negative
for attractive interaction) of a condensate consisting of $N$ particles with $s$-wave scattering
length $a_s$ and a dimensionless length unit $x_s$ \cite{Bao2010, Cai2010}. Furthermore
 $\Og\in\R$ is a given dimensionless constant representing
the angular velocity, and $\lambda=\frac{m N\mu_0\mu_{\rm dip}^2}{3\hbar^2 x_s}$ is a dimensionless constant
describing the strength of the long-range dipole-dipole interaction with $m$ the mass of a particle, $\mu_0$ the
vacuum permeability, $\mu_{\rm dip}$ the permanent magnetic dipole moment (e.g. $\mu_{\rm dip}=6\mu_{_B}$ for
$^{52}$C$_{\rm r}$ with $\mu_{_B}$ being the Bohr magneton) and $\hbar$ the Planck constant
\cite{Bao2010, Cai2010,Seiringer2003}.
In addition, $L_z$ is the dimensionless $z$-component of the angular momentum rotation defined as
\be\label{angular}
L_z = -i (x\p_y - y\p_x),
\ee
and $U_{\rm dip}$ is the dimensionless long-range dipole-dipole interaction potential defined as
\be\label{dipole}
U_{\rm dip}(\bx) = \fl{3}{4\pi|\bx|^3}\left[1-\fl{3(\bx\cdot\bn)^2}{|\bx|^2}\right]=
\fl{3}{4\pi|\bx|^3}\left[1-3\cos^2(\vartheta)\right], \qquad \bx\in{\Bbb R}^3,
\ee
with $\bn = (n_1, n_2, n_3)^T\in{\Bbb R}^3$ a given unit vector, i.e. $|\bn|=\sqrt{n_1^2+n_2^2+n_3^2}=1$,
representing  the dipole axis (or dipole moment) and $\vartheta=\vartheta_\bn(\bx)$
the angle between the dipole axis $\bn$ and
the vector $\bx$ \cite{Bao2010, Cai2010}. The wave function is normalized to
\be\label{norm478}
\|\psi\|^2:=\int_{{\mathbb R}^3} |\psi(\bx,t)|^2\,d\bx=1.
\ee
In addition, similar to \cite{Bao2010, Cai2010},
the above GPE (\ref{NGPE}) can be re-formulated as the following
Gross-Pitaevskii-Poisson system \cite{Bao2010,Bao20130, Cai2010}
\bea\label{GPPS1}
&&i\p_t \psi({\bx}, t) = \left[-\fl{1}{2}\nabla^2 + V({\bf x}) + (\kappa-\lambda) |\psi|^2 -
3\lambda \vphi(\bx,t) - \Omega L_{z}\right]\psi(\bx,t), \\
\label{GPPS1789}
&&\vphi(\bx,t)=\partial_{\bn\bn}u(\bx,t), \quad -\nabla^2u(\bx,t)=|\psi(\bx,t)|^2 \ \
\hbox{with}\ \  \lim_{|\bx|\to\infty} u(\bx,t)=0,\label{GPPS2}
\eea
where $\partial_\bn=\bn\cdot \nabla$ and $\partial_{\bn\bn}=\partial_\bn(\partial_\bn)$. From (\ref{GPPS2}),
it is easy to see that
\be\label{3Dconv}
u(\bx, t) = \left(\fl{1}{4\pi|\bx|}\right)*|\psi|^2:=\int_{{\mathbb R}^3}
\fl{1}{4\pi|\bx-\bx^\prime|} |\psi(\bx^\prime,t)|^2d\bx^\prime, \quad  \bx\in{\Bbb R}^3, \quad t\geq 0.
\ee

In some physical experiments of rotating BEC, the external trap is strongly confined in the $z$-direction, i.e.
\be\label{trapas}
V(\bx)=V_2(x,y)+\frac{z^2}{2\varepsilon^4}, \qquad \bx\in {\mathbb R}^3,
\ee
with $0<\varepsilon\ll1$ a given dimensionless parameter \cite{Bao20130}, resulting in a pancake-shaped BEC.
Similar to the case of a non-rotating BEC, formally the GPE (\ref{NGPE}) or (\ref{GPPS1})-(\ref{GPPS2}) can
effectively be approximated by a two-dimensional (2D) GPE as
\cite{Bao2006, Bao2010, Cai2010}:
\bea\label{GPPS3}
&&\qquad i\p_t \psi({\bx_\perp}, t) = \left[-\fl{1}{2}\nabla_\perp^2
+ V_2({\bf x}_\perp) + \frac{\kappa+\lambda(3 n_3^2-1)}{\varepsilon\sqrt{2\pi}} |\psi|^2 -\frac
{3\lambda}{2}\vphi
- \Omega L_{z}\right]\psi, \\
&&\qquad \vphi=\vphi(\bx_\perp,t)=\left(\partial_{\bn_\perp\bn_\perp}-
n_3^2\nabla_\perp^2\right)u(\bx_\perp,t),
\quad \bx_\perp=(x,y)^T\in {\mathbb R}^2,\ t\geq0,
\label{GPPS4}
\eea
where $\nabla_\perp=(\partial_x,\partial_y)^T$,
$\nabla_\perp^2=\partial_{xx}+\partial_{yy}$, $\bn_\perp=(n_1,n_2)^T$,
$\partial_{\bn_\perp}=\bn_\perp\cdot\nabla_\perp$, $\partial_{\bn_\perp\bn_\perp}=\partial_{\bn_\perp}(\partial_{\bn_\perp})$ and
\be\label{Uepsm}
u(\bx_\perp,t)=G^\varepsilon*|\psi|^2, \quad
G^\varepsilon(\bx_\perp)=\frac{1}{(2\pi)^{3/2}}\int_{\mathbb R}
\frac{e^{-s^2/2}}{\sqrt{|\bx_\perp|^2+\varepsilon^2s^2}}ds, \quad \bx_\perp\in{\mathbb R}^2.
\ee
The above problem (\ref{GPPS3})-(\ref{GPPS4}) with (\ref{Uepsm}) is usually called \emph{surface adiabatic model} (SAM)
for a rotating BEC with dipole-dipole interaction in 2D.
Furthermore, taking $\varepsilon\to0^+$ in (\ref{Uepsm}), we obtain
\be\label{kernl28}
G^\varepsilon(\bx_\perp)\to \frac{1}{2\pi|\bx_\perp|}:=G^0(\bx_\perp),
\qquad \bx_\perp\in{\mathbb R}^2.
\ee
This, together with (\ref{Uepsm}), implies that when $\varepsilon\to0^+$,
\be\label{Uepsm23}
u(\bx_\perp,t)=\frac{1}{2\pi|\bx_\perp|}*|\psi|^2\Longleftrightarrow (-\nabla_\perp^2)^{1/2}
u=|\psi|^2
\quad \hbox{with}\ \lim_{|\bx_\perp|\to\infty}u(\bx_\perp,t)=0.
\ee
The problem (\ref{GPPS3})-(\ref{GPPS4}) with (\ref{Uepsm23}) is usually called \emph{surface density model} (SDM)
for a rotating BEC with dipole-dipole interaction in 2D. Note that even for the SDM we retain the $\varepsilon$-dependence
in \eqref{GPPS3}.

In fact, the GPE (\ref{NGPE}) or (\ref{GPPS1}) in 3D and the SAM or SDM in 2D can be written
in a  unified way in $d$-dimensions ($d = 2$ or $3$) with $\bx=(x,y)^T$ when $d=2$ and $\bx=(x,y,z)^T$ when
$d=3$:
\bea\label{GGPE}
&&i\p_t \psi({\bx}, t) = \left[-\fl{1}{2}\nabla^2 + V({\bf x}) +
\bt|\psi|^2 +\eta \vphi(\bx,t)- \Omega L_{z}\right]\psi({\bx}, t), \\
\label{GGPE93}
&&\vphi(\bx,t)=L_\bn u(\bx,t), \qquad u(\bx,t)=G*|\psi|^2,
\qquad \bx\in{\mathbb R}^d, \quad t\geq0,
\eea
where $V(\bx)=V_2(\bx)$ when $d=2$ and
\be
\label{Gpotential}
\beta = \left\{\begin{array}{l}
\fl{\kappa +\lambda(3n_3^2-1)}{\varepsilon\sqrt{2\pi}}, \\
\kappa - \lambda,  \\
\end{array}\right. \quad \eta = \left\{\begin{array}{l}
-3\lambda/2, \\
-3\lambda,  \\
\end{array}\right. \quad L_\bn = \left\{\begin{array}{ll}
 \p_{\bf n_\bot n_\bot} - n_3^2\nabla^2, \quad & d = 2,\\
\p_{\bf nn}, & d = 3,
\end{array}\right.
\ee
\be\label{poten987}
G(\bx)=\left\{\begin{array}{l}
1/2\pi|\bx|, \\
G^\varepsilon(\bx), \\
1/4\pi|\bx|, \\
\end{array}\right.
\Longleftrightarrow \widehat{G}(\xi)=\left\{\begin{array}{ll}
1/|\xi|, &d=2 \ \&\ \hbox{SDM},\\
\frac{1}{2\pi^2}\int_{\mathbb R}\frac{e^{-\varepsilon^2 s^2/2}}{|\xi|^2+s^2}ds, &d=2\ \&\ \hbox{SAM},\\
1/|\xi|^2, &d=3,\\
\end{array}\right.
\ee
where $\widehat{f}(\xi)$ denotes the Fourier transform of the function $f(\bx)$ for $\bx,\,
\xi\in {\mathbb R}^d$. For studying the dynamics of a rotating BEC,
the following initial condition is used:
\be
\label{Ginitial}
\psi(\bx, 0) = \psi_0(\bx), \qquad \bx\in{\Bbb R}^d,\qquad\mbox{with}\quad
\|\psi_0\|^2 := \int_{\Bbb R^d}|\psi_0(\bx)|^2\,d\bx = 1.
\ee
We remark here that in most BEC experiments, the following dimensionless harmonic potential is used
\be\label{Vpoten}
V(\bx) = \fl{1}{2}\left\{\begin{array}{ll}
 \gm_x^2x^2 + \gm_y^2y^2, & d = 2,\\
\gm_x^2x^2 + \gm_y^2y^2 + \gm_z^2z^2, \ \ &d = 3,
\end{array}\right.
\ee
where $\gm_x>0$, $\gm_y>0$ and $\gm_z>0$ are  dimensionless constants proportional to the
trapping frequencies in $x$-, $y$- and $z$-direction, respectively.

Recently, many numerical and theoretical studies have been done on rotating (dipolar)
BECs.
There have been many numerical  methods proposed to study the dynamics of
non-rotating BECs, i.e. when $\Og = 0$ and $\eta = 0$ \cite{Cerimele2000, Aftalion2001, Kasamatsu2003, Bao2003, Bao2005, Tiwari0000,  Muruganandam2009}.  Among them,
the time-splitting sine/Fourier pseudospectral method is one of the most successful methods.
Compared to other methods, the time-splitting pseudospectral method has spectral accuracy in
space and is easy to implement. In addition, as was shown in \cite{Bao2010}, this method
can also be easily generalized to simulate the dynamics of dipolar BECs when $\eta \neq 0$.
However,  in rotating condensates,  i.e. when $\Og\neq0$,  we can not directly apply
the time-splitting pseudospectral method proposed in \cite{Bao2005} to study their dynamics due to the
appearance of angular rotational term.
So far, there have been several methods introduced to solve the GPE
with an angular momentum term. For example,  a pseudospectral type method was proposed
in \cite{Bao2006} by reformulating the problem in the
two-dimensional polar coordinates $(r,\theta)$ or three-dimensional cylindrical coordinates
$(r,\theta,z)$. The method is of second-order or fourth-order in the radial direction and spectral accuracy in other
directions.
A time-splitting alternating direction implicit  method was proposed in \cite{Bao2006-1},
where the authors decouple the angular terms into two parts and apply the Fourier transform in each direction.
Furthermore, a generalized Laguerre-Fourier-Hermite pseudospectral method was presented in
\cite{Bao2009}. These methods have higher spatial accuracy compared to those in
\cite{Aftalion2001, Kasamatsu2003,Bao2013-2} and are also valid in dissipative variants of the GPE (\ref{NGPE}), cf.\
\cite{Tsubota2002}. On the other hand, the implementation of these methods can become quite involved.
The aim of this paper is to propose a simple and efficient numerical method to solve the GPE with angular momentum rotation term which may include a dipolar interaction term.  One novel idea in this method consists
in the use of rotating Lagrangian coordinates as in \cite{Antonelli2012} in which the angular momentum rotation term vanishes.
Hence, we can easily apply the methods for non-rotating BECs in \cite{Bao2005} to solve the rotating case.

This paper is organized as follows.  In Section \ref{section2}, we present the dynamical laws
of rotating dipolar BECs based on the GPE (\ref{GGPE})--(\ref{Ginitial}).
Then in Section \ref{section3}, we introduce a coordinate transformation and cast
the GPE  and its dynamical quantities in the new coordinate system. Numerical methods are proposed
in Section \ref{section4} to discretize the GPE for both the two-dimensional and three-dimensional cases.
In Section \ref{section5}, we report on the accuracy of our methods and present some numerical
results. We make some concluding remarks in Section \ref{section6}.

\section{Dynamical properties}
\setcounter{equation}{0}
\label{section2}

In this section, we analytically study the dynamics of rotating dipolar BECs. We present dynamical laws, including the
conservation of angular momentum expectation, the dynamics of condensate widths and the dynamics of the center
of mass.  In the following we omit the proofs for brevity; they are similar to the ones in \cite{Bao2005, Bao2006}.


\subsection{Conservation of mass and energy}
\label{section2-1}
The GPE in (\ref{GGPE})--(\ref{Ginitial}) has two important invariants:  the
{\it mass} (or {\it normalization}) of the wave function,  which is defined as
\bea\label{norm}
N(t):=\|\psi(\cdot, t)\|^2 := \int_{{\Bbb R}^d}|\psi({\bx}, t)|^2 d{\bf x}
\equiv \int_{{\Bbb R}^d}|\psi({\bf x}, 0)|^2 d{\bf x} = 1, \qquad t\geq 0,
\eea
and the {\it energy  per particle}
\bea\label{energy}
E(t)&:=&E(\psi(\cdot, t))=\int_{{\Bbb R}^d}\left[\fl{1}{2}|\nabla\psi|^2 + V({\bf x})|\psi|^2+\fl{\bt}{2}
|\psi|^4+\fl{\eta}{2}\vphi |\psi|^2-\Og \psi^*L_z\psi\right]d{\bf x} \nn\\
&\equiv& E(\psi(\cdot, 0))=E(\psi_0), \qquad t\geq 0,
\eea
where $f^*$ denotes the conjugate of the complex-valued function $f$.
Stationary states, corresponding to critical points of the energy defined
 in (\ref{energy}), play an important role in
the study of rotating dipolar BECs.  Usually, to find
 stationary states $\phi_s(\bx)$, one can use the ansatz
\bea\label{Ansatz}
\psi(\bx, t) = e^{-i\mu_s t} \phi_s(\bx), \qquad \bx\in{\Bbb R}^d, \quad t\geq0,
\eea
where $\mu_s\in{\Bbb R}$ is the chemical potential.
Substituting (\ref{Ansatz}) into (\ref{GGPE}) yields the nonlinear eigenvalue problem
\bea\label{IGGPE}
&&\mu_s\phi_s({\bx}) = \left[-\fl{1}{2}\nabla^2 + V({\bf x}) + \bt|\phi_s|^2 +\eta \vphi_s
- \Omega L_{z}\right]\phi_s(\bx), \quad \
\bx\in{\Bbb R}^d, \\
\label{IGGPE97}
&&\vphi_s(\bx)=L_\bn u_s(\bx), \qquad u_s(\bx)=G*|\phi_s|^2,
\qquad \bx\in{\mathbb R}^d,
\eea
under the constraint
\bea\label{Inorm}
\|\phi_s\|^2 = \int_{\Bbb R^d}|\phi_s(\bx)|^2 d\bx = 1.
\eea
Thus, by solving the constrained nonlinear eigenvalue problem (\ref{IGGPE})--(\ref{Inorm}),  one can
find the stationary states of rotating dipolar BECs.  In physics literature, the stationary state
with the lowest energy is called {\it ground state}, while those with larger energy are called
{\it excited states}.

\subsection{Conservation of angular momentum expectation}
\label{section2-2}

The {\it angular momentum expectation} of a condensate is defined as \cite{Bao2005, Bao2006}
\bea\label{Def-Angular}
\langle L_z\rangle(t) =  \int_{{\Bbb R}^d} \psi^*(\bx,t)  L_z \psi(\bx,t)\,d\bx,\qquad t\geq 0.
\eea
This quantity is often used to measure the  vortex flux.  The following lemma describes  the dynamics of angular
momentum expectation in rotating dipolar BECs.

\begin{lemma}\label{lemma1}
Suppose that $\psi(\bx,t)$ solves the GPE (\ref{GGPE})--(\ref{Ginitial}) with $V(\bx)$ chosen
as the harmonic potential (\ref{Vpoten}).  Then we have
\bea\label{Eq0-Lemma1}
\qquad \fl{d\langle L_z\rangle(t)}{dt} = (\gamma_x^2-\gamma_y^2) \int_{{\Bbb R}^d} x y |\psi|^2 d\bx
-\eta { \int_{{\Bbb R}^d} |\psi|^2
\left[(x\p_y-y\p_x) \varphi\right]d\bx},\quad t\geq 0.
\eea
Furthermore, if the following two conditions are satisfied: (i) $\gamma_x = \gamma_y$
and (ii) $\eta = 0$ for any given initial data $\psi_0$ in (\ref{Ginitial})
or  $\bn = (0, 0, 1)^T$ when $\psi_0$ satisfies $\psi_0(\bx)=f(r) e^{im\theta}$ in 2D and
$\psi_0(\bx)=f(r,z) e^{im\theta}$ in 3D with ${m\in \mathbb Z}$ and
$f$ a function of $r$ in 2D or $(r,z)$ in 3D, then the angular momentum expectation
is conserved, i.e.\
\bea\label{Eq2-Lemma1}
\langle L_z\rangle(t) \equiv \langle L_z \rangle(0), \quad t\geq0.
\eea
That is, in a radially symmetric trap in 2D or a cylindrically symmetric
trap in 3D, the angular momentum expectation is
conserved when either there is no dipolar interaction for any initial data
or the dipole axis is parallel to the $z$-axis with a radially/cylindrically
symmetric or central  vortex-type initial data.
\end{lemma}

\subsection{Dynamics of condensate width}
\label{section2-3}

The {\it condensate width} of a BEC in $\ap$-direction (where $\ap = x, y, z$ or $r=\sqrt{x^2+y^2}$) is defined by
\bea\label{Def-Width}
\sigma_\ap(t) = \sqrt{\delta_\ap(t)}, \quad \  t\geq 0, \qquad \mbox{where}\quad\dt_{\ap}(t)
= \int_{{\Bbb R}^d}\ap^2|\psi(\bx,t)|^2 d\bx.
\eea
In particular, when $d = 2$, we have the following lemma for its dynamics \cite{Bao2006}:

\begin{lemma}\label{lemma2} Consider two-dimensional BECs with radially symmetric harmonic trap
(\ref{Vpoten}),  i.e. $d
= 2$ and $\gamma_x = \gm_y : = \gm_r$. If $\eta = 0$,
then for any initial datum $\psi_0(\bx)$ in (\ref{Ginitial}), it holds for $t\ge0$
\bea\label{solution-ODE}
\qquad \dt_r (t) = \fl{E(\psi_0)+\Og\langle L_z\rangle(0)}{\gm_r^2}[1-\cos(2\gamma_rt)] + \dt_r^{(0)}\cos(2\gm_rt)
+\fl{\dt_r^{(1)}}{2\gm_r}\sin(2\gm_rt),
\eea
 where $\dt_r(t) := \dt_x(t) + \dt_y(t)$, $\dt_r^{(0)} := \dt_x(0)
 +\dt_y(0)$ and $\dt_r^{(1)}  := \dot{\dt}_x(0) + \dot{\dt}_y(0)$. Furthermore,
 if the initial condition $\psi_0(\bx)$ is {radially symmetric}, we have
\bea \label{dtx9845}
\dt_x(t) &=& \dt_y(t) \;=\; \fl{1}{2}\dt_r (t)\nonumber\\
&=& \fl{E(\psi_0)+\Og\langle L_z\rangle(0)}{2\gm_x^2}[1-\cos(2\gamma_xt)] + \dt_x^{(0)}\cos(2\gm_xt)
+\fl{\dt_x^{(1)}}{2\gm_x}\sin(2\gm_xt), \ t\ge0.
\eea
Thus, in this case the condensate widths $\sigma_x(t)$ and $\sigma_y(t)$ are periodic functions
with frequency doubling trapping frequency.

\end{lemma}

\subsection{Dynamics of center of mass}
\label{section2-4}

We define the {\it center of mass} of a condensate at any time $t$ by
\bea\label{Def-Cmass}
\bx_c(t) = \int_{\Bbb R^d} \bx \, |\psi(\bx,t)|^2 d\bx, \qquad t\geq0.
\eea
The following lemma describes the dynamics of the center of mass.
\begin{lemma}\label{lemma4}
Suppose that $\psi(\bx,t)$ solves the GPE (\ref{GGPE})--(\ref{Ginitial}) with $V(\bx)$ chosen
as the harmonic potential (\ref{Vpoten}). Then for any given initial data $\psi_0$, the dynamics
of the center of mass are governed by the following second-order ODEs:
\bea
\label{ODE2}
&&\ddot{\bx}_c(t) - 2\Og{J} \dot{\bx}_c(t)+ ({ \Lambda}+\Og^2{ J^2})\, \bx_c(t) = 0, \qquad t\geq 0,
\qquad\qquad\qquad\\
\label{ODE0}
&& \bx_c(0) = \bx_c^{(0)} :=\int_{\Bbb R^d} \bx|\psi_0|^2 d\bx, \qquad\\
\label{ODE1}
&&\dot{\bx}_c(0) = \bx_c^{(1)} : = \int_{\Bbb R^d}{\rm Im}\left(\psi_0^*\nabla\psi_0\right)d\bx
-\Og{ J} \bx_c^{(0)},
\eea
where ${\rm Im}(f)$ denotes the imaginary part of the function $f$ and the matrices
\beas
{J} = \left(\begin{array}{cc}
0 & 1  \\
-1 & 0
\end{array}\right), \qquad
{\Lambda} = \left(\begin{array}{cc}
\gm_x^2 & 0 \\
0 & \gm^2_y
 \end{array}\right), \qquad
\mbox{for\quad $d = 2$},
\eeas
or
\beas
{J} = \left(\begin{array}{ccc}
0 & 1 & 0 \\
-1 & 0 & 0 \\
0 & 0 & 0
\end{array}\right), \qquad
{\Lambda} = \left(\begin{array}{ccc}
\gm_x^2 & 0 & 0 \\
0 & \gm^2_y & 0 \\
0 & 0 & \gm_z^2
\end{array}\right), \qquad
\mbox{for\quad $d = 3$}.
\eeas
\end{lemma}

Lemma \ref{lemma4} shows that the dynamics of the center of mass depends on the trapping frequencies and
the  angular velocity, but it is independent of the interaction strength constants $\beta$ and $\eta$
in (\ref{GGPE}). For analytical solutions to the second-order ODEs (\ref{ODE2})-(\ref{ODE1}),
we refer to \cite{Zhang2007}.

\subsection{An analytical solution under special initial data}

 From  Lemma \ref{lemma4}, we can construct an analytical solution to the GPE
(\ref{GGPE})--(\ref{Ginitial}) when the initial data is chosen as a stationary state
with its center shifted.

\begin{lemma}
\label{lemma5} Suppose $V(\bx)$ in (\ref{GGPE}) is chosen as the harmonic
potential (\ref{Vpoten}) and the initial condition $\psi_0(\bx)$ in (\ref{Ginitial}) is chosen as
\bea\label{Eq0-Lemma24}
\psi_0(\bx) = \phi_s(\bx - \bx^0), \qquad \bx\in{\Bbb R}^d,
\eea
where $\bx^0\in{\Bbb R}^d$ is a given  point and $\phi_s(\bx)$ is a stationary state defined in (\ref{IGGPE})--(\ref{Inorm}) with chemical potential $\mu_s$,  then the exact solution of
(\ref{GGPE})--(\ref{Ginitial}) can be constructed as
\bea
\psi(\bx, t) = \phi_s(\bx-\bx_c(t))\,e^{-i\mu_s t }\,e^{i w(\bx,t)}, \qquad \bx\in{\Bbb R}^d, \quad t\ge0,
\eea
where  $\bx_c(t)$ satisfies the ODE (\ref{ODE2}) with
\bea
\bx_c(0) =  \bx^0, \qquad \dot{\bx}_c(0) = -\Og{J}\bx^0,
\eea
and
$w(\bx, t)$ is linear in $\bx$, i.e.\
\beas
w(\bx,t) = {\bf c}(t)\cdot \bx  + g(t), \quad {\bf c}(t) = (c_1(t), \ldots, c_d(t))^T, \quad \bx\in{\Bbb R}^d, \ \
 t\geq0
\eeas
for some functions ${\bf c}(t)$ and  $g(t)$. Thus, up to phase shifts, $\psi$ remains a stationary state with shifted center
at all times.
\end{lemma}

\section{GPE under a rotating Lagrangian coordinate}
\setcounter{equation}{0}
\label{section3}

In this section, we first introduce a coordinate transformation and derive the GPE in transformed coordinates.
Then we reformulate the dynamical quantities studied in Section \ref{section2} in the new coordinate system.

\subsection{A rotating Lagrangian coordinate transformation}
\label{section3-1}

\begin{figure}[h!]
\centerline{
\psfig{figure=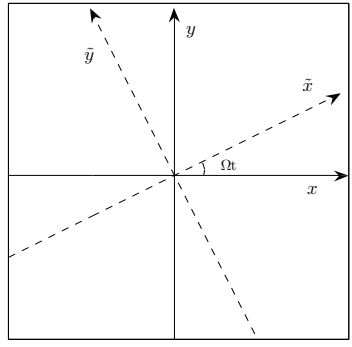,height=5.5cm,width=7.4cm,angle=0}}
\caption{Cartesian (or Eulerian) coordinates $(x,y)$ (solid) and rotating Lagrangian
coordinates $(\tilde{x}, \tilde{y})$ (dashed) in 2D for any fixed $t\ge0$. }\label{rot-axis}
\end{figure}

For any time $t\geq 0$, let ${ A}(t)$ be an orthogonal rotational matrix  defined as
\bea\label{Amatrix}
{ A}(t)=\left(\begin{array}{cc}
\cos(\Omega t) & \sin(\Omega t) \\
-\sin(\Omega t) & \cos(\Omega t)
 \end{array}\right),  \quad\ \  \mbox{if \ \ $d = 2$,} \qquad \quad  \ \
         \eea
         and
         \bea\label{Matrixa}
{A}(t)=\left(\begin{array}{ccc}
         \cos(\Omega t) & \sin(\Omega t) & 0 \\
         -\sin(\Omega t) & \cos(\Omega t) & 0 \\
         0 & 0  & 1
         \end{array}\right), \quad\ \  \mbox{if \ \ $d = 3$.} \qquad
\eea
It is easy to verify that $A^{-1}(t) =A^T(t)$ for any $t\ge0$ and ${A}(0) = { I}$ with
 $I$ the identity matrix. For any $t\ge0$, we introduce the {\it rotating Lagrangian coordinates}
$\tbx$ as \cite{Antonelli2012,Garcia2001,Hintermueller2012}
\bea\label{transform}
\tbx={A}^{-1}(t) \bx=A^T(t)\bx \quad \Leftrightarrow \quad \bx= {A}(t){\tbx},   \qquad \bx\in {\mathbb R}^d,
\eea
and denote the wave function in the new coordinates as $\phi:=\phi(\tbx, t)$
\bea\label{transform79}
\phi(\tbx, t):=\psi(\bx, t)= \psi\left({A}(t){\tbx},t\right), \qquad \bx\in {\mathbb R}^d, \quad t\geq0.
\eea
In fact, here we refer the Cartesian coordinates $\bx$ as the {\it Eulerian coordinates} and
Fig. \ref{rot-axis} depicts the geometrical relation between the Eulerian coordinates $\bx$
and the rotating Lagrangian coordinates $\tbx$ for any fixed $t\ge0$.

Using the chain rule, we  obtain the derivatives
\beas\label{operator1}
&&\qquad \p_t\phi(\tbx,t) =\p_t\psi(\bx, t)  + \nabla\psi(\bx,t)\cdot\left(\dot{A}(t) \tbx\right) = \p_t\psi(\bx,t)
- \Og(x\p_y-y\p_x)\psi(\bx,t),\\
\label{operator2}
&&\qquad \nabla\phi(\tbx,t) = {A^{-1}}(t)\nabla\psi(\bx,t) , \qquad \nabla^2\phi(\tbx, t) = \nabla^2\psi(\bx,t).
\eeas
 Substituting them  into (\ref{GGPE})--(\ref{Ginitial}) gives  the  following
$d$-dimensional GPE in the rotating Lagrangian coordinates:
\bea\label{GPE}
&&\qquad i\p_t \phi(\widetilde{\bx}, t) = \left[-\fl{1}{2}\nabla^2 + W(\widetilde{\bx}, t) + \bt|\phi|^2
+\eta \vphi(\tbx,t) \right]\phi(\widetilde{\bf x},t), \quad \tbx\in{\mathbb R}^d, \quad t>0,\\
\label{varphi}
&&\qquad \vphi(\tbx,t)=L_{{\bf m}(t)}u(\tbx,t), \qquad u(\tbx,t)=G*|\phi|^2,\qquad \tbx\in{\mathbb R}^d, \quad t\geq0,
\eea
where $G$ is defined in \eqref{poten987} and
\bea\label{wtbx534}
&&W(\tbx,t) = V(A(t)\tbx), \qquad \tbx\in {\mathbb R}^d,\\
\label{lmt8765}
&&L_{{\bf m}(t)} = \left\{\begin{array}{ll}
\p_{{\bf m}_\perp(t) {\bf m}_\perp(t)} - n_3^2\nabla^2, \quad &d = 2,\\
\p_{{\bf m}(t) {\bf m}(t)}, &d = 3,
\end{array}\right.\quad  t\geq0,
\eea
with ${\bf m}(t) \in {\mathbb R}^3$ and ${\bf m}_\perp(t) \in {\mathbb R}^2$ defined as
\be\label{Axis}
{\bf m}(t) =  \left(\begin{array}{c}
m_1(t) \\
m_2(t) \\
m_3(t) \end{array}\right) : = {A}^{-1}(t)\bn =  \left(\begin{array}{c}
n_1\cos(\Og t) - n_2\sin(\Og t)\\
n_1\sin(\Og t) + n_2\cos(\Og t)\\
n_3
\end{array}\right), \quad t\geq0,
\ee
and ${\bf m}_\perp(t)=(m_1(t),m_2(t))^T$, respectively. The initial data transforms as
\be
\label{Initial}
\phi(\tbx, 0) = \psi(\bx,0)=\psi_0(\bx):=\phi_0(\bx)=\phi_0(\tbx), \qquad \tbx=\bx\in{\Bbb R}^d.
\ee
We remark here again that if $V(\bx)$ in (\ref{GGPE})  is a harmonic  potential as defined in (\ref{Vpoten}),
then the potential $W(\tbx, t)$ in (\ref{GPE}) has the form
\beas
W(\tbx, t) = \fl{1}{4}\left\{\begin{array}{ll}
{\og}_1(\tx^2 + \ty^2) + {\og}_2\left[(\tx^2-\ty^2)\cos(2\Og t) + 2\tx\ty\sin(2\Og t)\right],
& d= 2, \\
\og_1(\tx^2 + \ty^2) + \og_2\left[(\tx^2-\ty^2)\cos(2\Og t) + 2\tx\ty\sin(2\Og t)\right]+2\gm_z^2\tz^2,
& d= 3, \\
\end{array}\right.
\eeas
where $\og_1 = \gm_x^2+\gm_y^2$ and $\og_2 = \gm_x^2 - \gm_y^2$. It is easy to see that when
$\gm_x = \gm_y := \gm_r$, i.e. radially and  cylindrically symmetric harmonic trap in 2D and
 3D, respectively,  we have $\og_1 = 2\gm_r^2$ and $\og_2 = 0$ and thus the potential
$W(\tbx, t) = V(\tbx)$ becomes time-independent.

In contrast to (\ref{GGPE}),  the GPE (\ref{GPE})  does not have an angular momentum rotation term, which
enables us to develop simple and efficient numerical methods for simulating the dynamics of rotating dipolar
BEC in Section \ref{section4}.
\subsection{Dynamical quantities}
\label{section3-2}

In the above, we introduced rotating Lagrangian coordinates and cast the GPE in the new coordinate system.
Next we consider the dynamical laws in terms of the new wave function $\phi({\widetilde{\bf x}}, t)$.

\bigskip
\noindent{\bf Mass and energy. }  In rotating Lagrangian coordinates, the conservation of mass \eqref{norm} yields
\bea\label{Norm}
\qquad \|\psi(\cdot, t)\|^2 :=  \int_{{\Bbb R}^{d}}|\psi(\bx,t)|^2d\bx =\int_{\Bbb R^d} |\phi(\tbx,t)|^2 d\tbx
= \|\phi(\cdot, t)\|^2 \equiv 1, \quad \ t\geq 0.
\eea
The energy  defined in \eqref{energy} becomes
\bea\label{Energy}
&&\widetilde{E}(\phi(\cdot, t))= \int_{{\Bbb R}^d}\left[\fl{1}{2}|\nabla\phi|^2 + W(\tbx,t)|\phi|^2+\fl{\bt}{2}
|\phi|^4+\fl{\eta}{2}\vphi |\phi|^2\right]d\tbx \nn\\
&&\qquad\qquad -\int_{{\Bbb R}^d} \int_0^t \Big[\partial_\tau W(\tbx,\tau)
+ \fl{\eta}{2}(\partial_\tau L_{{\bf m}(\tau)})
u(\tbx, \tau)\Big] |\phi|^2 d\tau\,d\tbx \equiv   \widetilde{E}(\phi(\cdot, 0)), \quad t\geq0,
\eea
where $u$ is given in \eqref{varphi}. Specifically, it holds
\[
 \partial_t L_{{\bf m}(t)} = 2\begin{cases}
                               \partial_{\dot{A}^T(t) {\bf n}_\perp} \partial_{A^T(t) {\bf n}_\perp}, \qquad &d=2,\\
                               \partial_{\dot{A}^T(t) {\bf n}} \partial_{A^T(t) {\bf n}}, \quad &d=3.\\
                             \end{cases}
\]

\bigskip
\noindent{\bf Angular momentum expectation. }
The angular  momentum expectation in the new coordinates becomes
\bea\label{AME}
\langle L_z\rangle(t)   &=&
-i\int_{\Bbb R^d} \psi^*(\bx,t)(x\p_y - y\p_x)\psi(\bx,t)\, d\bx \nn\\ &=&
 -i\int_{\Bbb R^d} \phi^*(\tbx,t)(\tx\p_{\tilde y} - \ty\p_{\tilde x})\phi(\tbx,t)\, d\tbx,\quad \  t\geq0,
\eea
which has the same form as (\ref{Def-Angular}) in the new
coordinates of $\tbx\in{\Bbb R^d}$ and the  wave function $\phi(\tbx, t)$. Indeed, if
we denote $\widetilde{L}_z$ as the $z$-component
of the angular momentum in the rotating Lagrangian coordinates,   we have
$\widetilde{L}_z = -i(\tx\p_{\tilde y} - \ty\p_{\tilde x}) = -i(x\p_y - y\p_x) = L_z$, i.e.\ the coordinate transform
does not
change the angular momentum in $z$-direction. In addition, noticing that for any $t\geq 0$ it holds
$\phi(\tbx, t) = \psi(\bx, t)$ and  $|{A}(t)| = 1$ for any $t\ge0$ immediately yields (\ref{AME}).

\bigskip
\noindent{\bf Condensate width. } After the coordinate transform, it holds
\bea
&&\dt_r(t) = \int_{\Bbb R^d} (x^2+y^2)|\psi|^2 d\bx =
\int_{\Bbb R^d} (\tx^2+\ty^2)|\phi|^2 d\tbx = \dt_{\tilde{x}}(t) + \dt_{\tilde{y}}(t), \\
&&\dt_z(t) = \int_{\Bbb R^d} z^2|\psi|^2 d\bx = \int_{\Bbb R^d} \tz^2 |\phi|^2 d\tbx  = \dt_{\tilde{z}}(t),
\eea
for any $t\geq 0$.

\bigskip
\noindent{\bf Center of mass.} The center of mass in rotating Lagrangian coordinates is defined as
\begin{equation}\label{DefCenterMass}
\tbx_c(t) =  \int_{\Bbb R^d} \tbx\, |\phi(\tbx, t)|^2 d\tbx , \qquad t\geq0.
\end{equation}
Since $\det(A(t)) = 1$ for any $t\ge0$, it holds that  $\bx_c(t) = A(t)\tbx_c(t)$ for any time $t \geq0$. In rotating Lagrangian coordinates,
we have the following analogue of Lemma  \ref{lemma5}:
\begin{lemma} Suppose $V(\bx)$ in (\ref{GGPE}) is chosen as the harmonic
potential (\ref{Vpoten}) and the initial condition $\phi_0(\tbx)$ in (\ref{Initial}) is chosen as
\bea
\phi_0(\tbx) = {\phi_s}(\tbx - \tbx^0), \qquad \tbx\in{\mathbb R}^d,
\eea
where $\tbx^0$ is a given point in ${\Bbb
R}^d$ and ${\phi_s}(\tbx)$ is a stationary state defined in
(\ref{IGGPE})--(\ref{Inorm}) with chemical potential ${\mu_s}$,
then the exact solution of \eqref{GPE}--\eqref{varphi} is of the form
\bea
\phi(\tbx, t)
&=& {\phi_s}(\tbx-\tbx_c(t))\,e^{-i{\mu_s} t }\,e^{i \widetilde{w}(\tbx,t)}, \qquad t > 0,
\eea
where  $\tbx_c(t)$ satisfies the second-order ODEs:
\bea
\label{xbtt345}
&&\ddot{\tbx}_c(t)  + A^T(t)\Lambda A(t)\, \tbx_c(t) = {\bf 0}, \qquad t\geq 0,
\qquad\qquad\qquad\\
\label{xbtt346}
&& \tbx_c(0) = \tbx^{0},\quad \dot{\tbx}_c(0)= {\bf0},
\eea
with the matrix $\Lambda$  defined in Lemma \ref{lemma4} and
$\widetilde{w}(\tbx, t)$ is linear in $\tbx$, i.e.\
\[
\widetilde{w}(\tbx,t) = {\bf \widetilde{c}}(t)\cdot \tbx  + \widetilde{g}(t), \quad {\bf \widetilde{c}}(t) =
(\widetilde{c}_1(t), \ldots, \widetilde{c}_d(t))^T, \quad \bx\in{\Bbb R}^d, \ \
 t\geq0.
\]
\end{lemma}

We have seen that the form of the transformation matrix ${A}(t)$ in (\ref{Matrixa}) is such that the coordinate transformation does not affect the quantities in $z$-direction, e.g.\ $\langle L_z\rangle$, $\sigma_z(t)$ and $z_c(t)$.

\section{Numerical methods}
\setcounter{equation}{0}
\label{section4}

To study the dynamics of rotating dipolar BECs, in this section we propose a simple and efficient
numerical method for discretizing the GPE (\ref{GPE})--(\ref{Initial})
 in rotating Lagrangian coordinates.  The detailed
discretizations for both the 2D and 3D GPEs are presented. Here we assume $\Omega\ne0$, and
for $\Omega=0$, we refer to  \cite{Bao2010,Bao2013,Xiong2009,Cai2010}.

In practical computations, we first truncate the whole space problem (\ref{GPE})--(\ref{Initial}) to a
bounded computational domain $\mathcal{D}\subset{\Bbb R}^d$ and consider
\bea\label{CGPE}
&&\qquad i\p_t\phi(\tbx, t) = -\fl{1}{2}\nabla^2\phi + W(\tbx, t)\phi + \beta |\phi|^2 \phi +\eta \varphi \phi, \quad
\tbx\in{\mathcal D},\ t > 0,\\
\label{Cvarphi}
&&\qquad \vphi(\tbx,t) = L_{{\bf m}(t)}u(\tbx,t), \quad u(\tbx,t)=\int_{{\Bbb R}^d}G(\tbx-\widetilde{\bf
y})\rho(\widetilde{\bf y},t) \,d\widetilde{\bf y},\quad \tbx\in{\mathcal D}, \ t>0;
\eea
where
\[\rho(\widetilde{\bf y},t)=\left\{\ba{ll}
|\phi(\widetilde{\bf y},t)|^2, &\widetilde{\bf y}\in \mathcal{D},\\
0, &\hbox{otherwise},
\ea\right. \qquad  \widetilde{\bf y}\in {\Bbb R}^d.
\]
The initial condition is given by
\bea
\label{Cinitial}
\phi(\tbx, 0) = \phi_0(\tbx), \quad \  \tbx\in \overline{\mathcal D}.
\eea
The boundary condition to (\ref{CGPE}) will be chosen based on the kernel function
$G$  defined in \eqref{poten987}. Due to the convolution in (\ref{Cvarphi}),
it is natural to consider using the Fourier transform to compute $u(\tbx,t)$.
However, from \eqref{poten987} and (\ref{Norm}), we know that
$\lim_{\xi\to0}\widehat{G}(\xi)=\infty$ and $\widehat{|\phi|^2}(\xi=0)\ne0$.
As noted for simulating dipolar BECs in 3D \cite{Ronen2006,Blakie2009,Bao2010}, there is a numerical locking phenomena,
i.e. numerical errors will be bounded below no matter how small the mesh size is, when
one uses the Fourier transform to evaluate $u(\tbx,t)$ and/or $\vphi(\tbx,t)$ numerically in
(\ref{Cvarphi}). As noticed in \cite{Bao2010,Bao20130}, the second (integral) equation in (\ref{Cvarphi}) can be
reformulated into the Poisson equation (\ref{GPPS1789}) and square-root-Poisson equation
(\ref{Uepsm23}) for 3D and 2D SDM model, respectively. With these PDE formulations for $u(\tbx,t)$,
we can truncate them on the domain $\mathcal D$ and solve them numerically via spectral method
with sine basis functions instead of Fourier basis functions and thus we can avoid using
the $0$-modes \cite{Bao2010}. Thus in 3D and 2D SDM model, we
choose the  homogeneous Dirichlet boundary conditions to (\ref{CGPE}). Of course,
for the 2D SAM model, one has to use the Fourier transform to compute $u(\tbx,t)$,
thus we take the  periodic boundary conditions  to (\ref{CGPE}).

\begin{figure}[h!]
\centerline{\psfig{figure=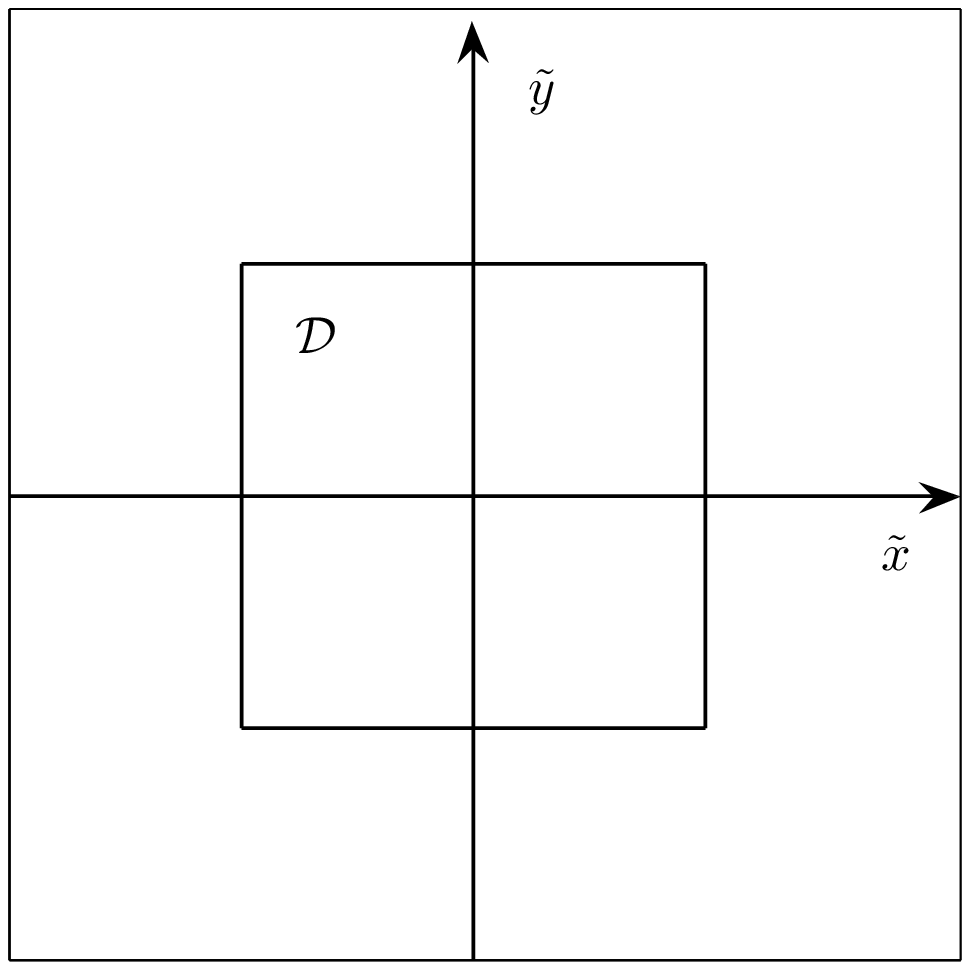,height=4.8cm,width=6.3cm,angle=0}
\psfig{figure=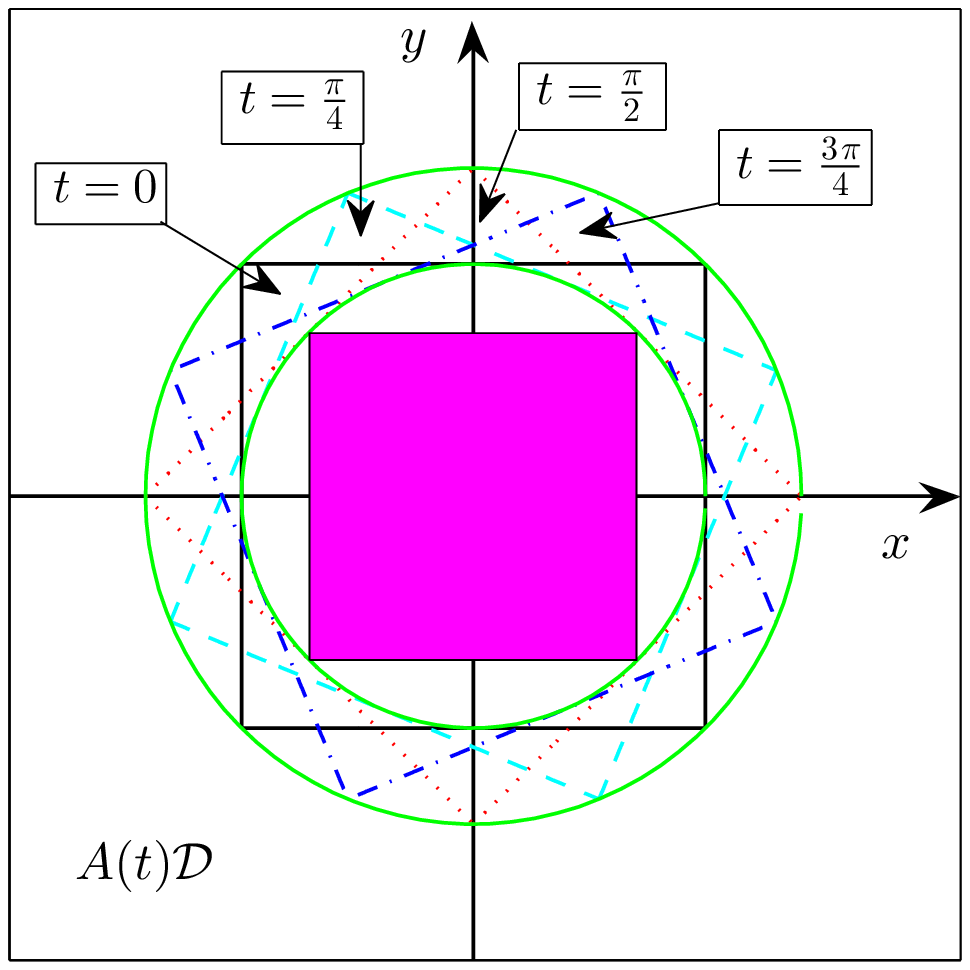,height=4.8cm,width=6.3cm,angle=0}}
\caption{The bounded computational domain $\mathcal{D}$ in rotating Lagrangian coordinates $\tbx$ (left) and
the corresponding domain $A(t)\mathcal{D}$ in Cartesian (or Eulerian) coordinates $\bx$ (right) when $\Omega=0.5$
at different times: $t=0$ (black solid), $t=\frac{\pi}{4}$ (cyan dashed), $t=\frac{\pi}{2}$ (red dotted)
and $t=\frac{3\pi}{4}$ (blue dash-dotted).  The two green solid circles determine two disks which
are the union (inner circle) and the intersection of all domains $A(t)\mathcal{D}$ for $t\geq0$, respectively. The magenta area is the vertical maximal square inside the inner circle.}\label{BackToOri}
\end{figure}

The computational domain ${\mathcal D}\subset{\Bbb R}^d$ is chosen as
${\mathcal D} = [a, b]\times[c, d]$ if $d=2$ and ${\mathcal D} = [a, b]\times[c, d]\times[e,f]$ if $d=3$.  Due to the
confinement of the external potential, the wave function decays exponentially fast as $|\tbx|\to \infty$.
 Thus if we choose ${\mathcal D}$ to be sufficiently large, the error from the domain truncation can be neglected.
As long as we solve $\phi(\tbx,t)$ in the bounded computational domain ${\mathcal D}$, we obtain a corresponding solution $\psi(\bx,t)$ in the domain $A(t){\mathcal D}$. As shown in Fig. \ref{BackToOri} for the example of a 2D domain, although the domains $A(t){\mathcal D}$ for $t \geq 0$, are in general different for different time $t$, they share a common disk which is bounded by the inner green solid
circle in Fig. \ref{BackToOri}. Thus, the value of $\psi(\bx,t)$ inside the vertical maximal square (the magenta area) which lies fully within the inner disk can be calculated easily by interpolation.

\subsection{Time-splitting method}
\label{section4-1}

Next, let us introduce a time-splitting method to discretize (\ref{CGPE})--(\ref{Cinitial}).
We choose a time-step size $\Dt t >0$ and define the time sequence as
$t_n  = n\Dt t$ for $n \in \mathbb{N}$.  Then from $t = t_n$ to $t = t_{n+1}$, we  numerically solve the GPE (\ref{CGPE}) in two steps. First we solve
\bea\label{1step}
i\p_t\phi(\tbx, t) &=& -\fl{1}{2}\nabla^2\phi(\tbx,t), \quad  \
\tbx\in{\mathcal D}, \quad t_n\leq t\leq t_{n+1}
\eea
for a time step of length $\Dt t$, and then we solve
\bea\label{2stepA}
&&i\p_t\phi(\tbx, t) = \left[W(\tbx, t)+ \beta|\phi|^2 + \eta \varphi\right]
\phi(\tbx,t),  \quad  \
\tbx\in{\mathcal D}, \quad t_n\leq t\leq t_{n+1},\\
&&\vphi(\tbx,t) = L_{{\bf m}(t)}u(\tbx,t), \qquad u(\tbx,t)=\int_{{\Bbb R}^d}G(\tbx-\widetilde{\bf y})
\rho(\widetilde{\bf y},t) \,d\widetilde{\bf y},
\label{2stepB}
\eea
for the same time step.

Equation (\ref{1step}) can be discretized in space by sine or Fourier pseudospectral methods
and  then integrated exactly in time.  If homogeneous Dirichlet  boundary conditions are used,
then we choose the sine pseudospectral method to discretize it; otherwise, the Fourier
pseudospectral method is used if the boundary conditions are periodic.  For more details, see
 e.g.\ \cite{Bao2005, Bao2010}.

On the other hand, we notice that on each time interval $[t_n, t_{n+1}]$, the problem
(\ref{2stepA})--(\ref{2stepB}) leaves
$|\phi(\tbx,t)|$ and hence $u(\tbx,t)$ invariant,  i.e.\ $|\phi(\tbx, t)| = |\phi(\tbx, t_n)|$  and
$u(\tbx, t) = u(\tbx, t_n)$ for all times $t_n\leq t \leq t_{n+1}$.
Thus, for $t\in[t_n, t_{n+1}]$,  Eq.\ (\ref{2stepA})  reduces  to
\bea\label{2stepA0}
&&\qquad i\p_t\phi(\tbx, t) = \left[W(\tbx, t)+ \beta|\phi(\tbx,t_n)|^2 + \eta\left(L_{{\bf m}(t)}u(\tbx, t_n)\right)\right]
\phi(\tbx,t),  \quad
\tbx\in{\mathcal D}.
\eea
Integrating (\ref{2stepA0}) in time gives the solution
\bea\label{2solution}
\qquad \phi(\tbx,t) = \phi(\tbx, t_n)\exp\left[-i\left(\beta|\phi(\tbx,t_n)|^2(t-t_n)
+\eta\Phi(\tbx, t)+\int_{t_n}^t W(\tbx,\tau)d\tau
\right)\right]
\eea
for $\tbx\in{\mathcal D}$ and $t\in[t_n, t_{n+1}]$, where the function $\Phi(\tbx, t)$ is defined by
\bea\label{Phip}
\Phi(\tbx, t) = \int_{t_n}^t \left[L_{{\bf m}(\tau)}u(\tbx, t_n)\right] d\tau=
\left(\int_{t_n}^t L_{\bf m(\tau)}\;d\tau \right) u(\tbx, t_n).
\eea
Plugging (\ref{Axis}) and (\ref{lmt8765}) into (\ref{Phip}), we get
\bea\label{Phip956}
\Phi(\tbx, t) = \tilde{L}_d(t) u(\tbx, t_n), \qquad \tbx\in {\mathcal D}, \quad t_n\le t\le t_{n+1},
\eea
where
\[
\tilde{L}_d(t)=\left\{\ba{ll}
[l_e^{11}(t)-l_e^{33}(t)]\p_{\tilde{x}\tilde{x}}+
[l_e^{22}(t)-l_e^{33}(t)]\p_{\tilde{y}\tilde{y}}+l_e^{12}(t)\p_{\tilde{x}\tilde{y}}, &d=2,\\
l_e^{11}(t)\p_{\tilde{x}\tilde{x}}+l_e^{22}(t)\p_{\tilde{y}\tilde{y}}
+l_e^{33}(t)\p_{\tilde{z}\tilde{z}}+l_e^{12}(t)\p_{\tilde{x}\tilde{y}}
+l_e^{13}(t)\p_{\tilde{x}\tilde{z}}+l_e^{23}(t)\p_{\tilde{y}\tilde{z}},   &d=3,
 \ea\right.
\]
with
\beas 
l_e^{11}(t)&=&\int_{t_n}^t m_1^2(\tau)d\tau=\int_{t_n}^t \left[n_1^2\cos^2(\Omega \tau)+n_2^2\sin^2(\Omega \tau)
-n_1n_2\sin(2\Omega \tau)\right]d\tau  \\
&=&\frac{n_1^2+n_2^2}{2}(t-t_n)+\frac{n_1^2-n_2^2}{4\Omega}\left[\sin(2\Omega t)-\sin(2\Omega t_n)\right]
+\frac{n_1n_2}{2\Omega}\left[\cos(2\Omega t)-\cos(2\Omega t_n)\right],
\eeas
\beas
l_e^{22}(t)&=&\int_{t_n}^t m_2^2(\tau)d\tau=\int_{t_n}^t \left[n_2^2\cos^2(\Omega \tau)+n_1^2\sin^2(\Omega \tau)
+n_1n_2\sin(2\Omega \tau)\right]d\tau  \\
&=&\frac{n_1^2+n_2^2}{2}(t-t_n)-\frac{n_1^2-n_2^2}{4\Omega}\left[\sin(2\Omega t)-\sin(2\Omega t_n)\right]
-\frac{n_1n_2}{2\Omega}\left[\cos(2\Omega t)-\cos(2\Omega t_n)\right],
\eeas  
\beas
l_e^{12}(t)&=&2\int_{t_n}^t m_1(\tau)m_2(\tau)d\tau=\int_{t_n}^t\left[(n_1^2-n_2^2)\sin(2\Omega \tau)+2n_1n_2\cos(2\Omega t)\right]d\tau\\
&=&\frac{n_1^2-n_2^2}{2\Omega}
\left[\cos(2\Omega t_n)-\cos(2\Omega t)\right]+\frac{n_1n_2}{\Omega}\left[\sin(2\Omega t)-\sin(2\Omega t_n)\right],
\eeas
\beas
l_e^{13}(t)&=&2n_3\int_{t_n}^t m_1(\tau)d\tau=2n_3 \int_{t_n}^t
\left[n_1\cos(\Omega \tau)-n_2\sin(\Omega \tau)\right]d\tau\\
&=&\frac{2n_3}{\Omega}
\left[n_1\left[\sin(\Omega t)-\sin(\Omega t_n)\right]+n_2\left[\cos(\Omega t)-\cos(\Omega t_n)\right]\right], \\
l_e^{23}(t)&=&2n_3\int_{t_n}^t m_2(\tau)d\tau=2n_3\int_{t_n}^t
\left[n_1\sin(\Omega \tau)+n_2\cos(\Omega \tau)\right]d\tau\\
&=&\frac{2n_3}{\Omega}
\left[n_2\left[\sin(\Omega t)-\sin(\Omega t_n)\right]-n_1\left[\cos(\Omega t)-\cos(\Omega t_n)\right]\right], \\
l_e^{33}(t)&=&\int_{t_n}^tn_3^2\, d\tau = n_3^2 (t-t_n), \qquad t_n\leq t\leq t_{n+1}.
\eeas
In Section \ref{section4-2}, we will discuss in detail the approximations to $\Phi(\tbx, t)$ in
(\ref{Phip956}).
In addition, we remark here again that if $V(\bx)$ in (\ref{GGPE})  is a harmonic  potential as defined in (\ref{Vpoten}), then the definite integral in (\ref{2solution}) can be calculated analytically as
\[\int_{t_n}^t W(\tbx,\tau)d\tau=\frac{1}{4}{\og}_1(\tx^2 + \ty^2) (t-t_n) +H(\tbx,t)+\frac{1}{2}
\left\{\ba{ll}
0, &d=2,\\
\gm_z^2\tz^2(t-t_n), &d=3,\\
\ea\right.
\]
where
\beas
H(\tbx,t)&=&\frac{1}{4}\int_{t_n}^t {\og}_2\left[(\tx^2-\ty^2)\cos(2\Og \tau)
+ 2\tx\ty\sin(2\Og \tau)\right] d\tau\\
&=&\frac{{\og}_2}{8\Omega}\left[\left(\tx^2-\ty^2\right)\left[\sin(2\Omega t)-\sin(2\Omega t_n)\right]
-2 \tx\ty \left[\cos(2\Omega t)-\cos(2\Omega t_n)\right]\right].
\eeas
Of course, for general external potential $V(\bx)$ in (\ref{GGPE}),
the integral of $W(\tbx, \tau)$ in (\ref{2solution}) might not be found analytically.
In this situation, we can simply adopt a  numerical quadrature to approximate it, e.g.
the Simpson's rule can be used as
\beas
\int_{t_n}^t W(\tbx, \tau)\,d\tau \approx \fl{t-t_n}{6}\left[W(\tbx, t_n) + 4W(\tbx, \fl{t_n+t}{2}) +
W(\tbx, t)\right].
\eeas

We remark here that, in practice, we always use the second-order Strang
 splitting method \cite{Strang} to combine the two steps in (\ref{1step})
and (\ref{2stepA})--\eqref{2stepB}. That is, from time $t=t_n$ to $t=t_{n+1}$, we (i)
 evolve (\ref{1step}) for half time step $\Dt t/2$ with initial data
 given at $t=t_n$; (ii) evolve (\ref{2stepA})--\eqref{2stepB}) for one step
 $\Dt t$ starting with the new data;
 and (iii) evolve (\ref{1step}) for half time step $\Dt t/2$ again with
 the newer data. For a more general discussion
of the splitting method, we refer the reader to \cite{Glowinski1989,Bao2013,Bao2005}.

\subsection{Computation of  $\Phi(\tbx, t)$}
\label{section4-2}

In this section, we present approximations to the function $\Phi(\tbx, t)$ in (\ref{Phip956}).
From the discussion in the previous subsection, we need only show how to discretize $u(\tbx,t_n)$
in (\ref{Cvarphi}) and its second-order derivatives in (\ref{Phip956}).

\subsubsection{Surface adiabatic model in 2D}
\label{section4-2-1}

In this case, the function $u(\tbx, t_n)$ in (\ref{Phip}) is given by
\bea\label{Un}
u(\tbx,t_n)=\int_{{\Bbb R}^2}G(\tbx-\widetilde{\bf y})
\rho(\widetilde{\bf y},t_n) \,d\widetilde{\bf y},
\quad \ \tbx\in{\mathcal D}.
\eea
with the kernel function $G$ defined in the second line of (\ref{poten987}). To approximate it,  we consider a
 2D box $\mathcal{D}$ with periodic boundary conditions.

Let $M$ and $K$ be  two even positive integers. Then we make the (approximate) ansatz
\bea\label{trans1276}
\qquad
u(\tbx, t_n) = \sum_{p=-M/2}^{M/2-1}\sum_{q=-K/2}^{K/2-1}
\widehat{u}^{f}_{pq}(t_n)\, e^{i\nu_p^1(\tx-a)}e^{i\nu_q^2(\ty-c)}, \qquad  \tbx=(\tx,\ty)\in {\mathcal D},
\eea
where $\widehat{u}^{f}_{pq}(t_n)$ is the Fourier coefficient of $u(\tbx, t_n)$ corresponding to the
frequencies $(p, q)$ and
\beas
\nu_p^1 = \fl{2p\pi}{b-a}, \qquad \nu_q^2 = \fl{2q\pi}{d-c}, \qquad (p, q)\in{\mathcal S}_{MK}.
\eeas
The index set ${\mathcal S}_{MK}$ is defined as
\beas
{\mathcal S}_{MK} =  \left\{(p, q)\,|\, -\fl{M}{2}\leq p\leq \fl{M}{2}-1, \ -\fl{K}{2}\leq q\leq \fl{K}{2}-1\right\}.
\eeas
We approximate the convolution in \eqref{Un} by a discrete convolution and take its discrete Fourier
transform to obtain
\bea\label{Fhat}
\widehat{u}^f_{pq}(t_n) = \widehat{G}(\nu_p^1,\nu_q^2)\cdot (\widehat{|\phi^n|^2})^f_{pq}, \qquad
(p, q)\in{\mathcal S}_{MK},
\eea
where $(\widehat{|\phi^n|^2})^f_{pq}$ is the Fourier coefficient corresponding to the frequencies $(p,q)$ of the
function $|\phi(\tbx, t_n)|^2$, and  $\widehat{G}(\nu_p^1,\nu_q^2)$ are given by (see details in (\ref{poten987}))
\bea\label{Uhat}
\widehat{G}(\nu_p^1,\nu_q^2) = \fl{1}{2\pi^2}\int_{-\infty}^{\infty}
\fl{e^{-\varepsilon^2 s^2/2}}{(\nu_p^1)^2
+ (\nu_q^2)^2+s^2}ds,\qquad (p, q)\in{\mathcal S}_{MK}.
\eea
Since the integrand in (\ref{Uhat}) decays exponentially fast, in practice we can first truncate it to an
 interval $[s_1, s_2]$ with $|s_1|, s_2 > 0$ sufficiently large  and then evaluate the truncated integral
by  using quadrature  rules, e.g. composite Simpson's or trapezoidal quadrature rule.

Combining (\ref{Phip956}), (\ref{trans1276}) and (\ref{Fhat}),  we obtain
an approximation of $\Phi(\tbx, t)$ in the solution (\ref{2solution}) via the Fourier spectral method as
\bea\label{phi765}\qquad\quad
\Phi(\tbx,t) = \sum_{p = -M/2}^{M/2-1}\sum_{q = -K/2}^{K/2-1}
\left[L(\nu_p^1, \nu_q^2, t)\widehat{G}(\nu_p^1,\nu_q^2)
\cdot (\widehat{|\phi^n|^2})^f_{pq}\right]
e^{i\nu_p^1(\tx-a)}e^{i\nu_q^2(\ty-c)},
\eea
for time $t_n \leq t\leq t_{n+1}$,  where the function $L(\xi_1, \xi_2, t)$ is defined as
\[ 
L(\xi_1, \xi_2, t)  =  -\left[\left(l_e^{11}(t)-l_e^{33}(t)\right)\xi_1^2 +
\left(l_e^{22}(t)-l_e^{33}(t)\right)\xi_2^2 +
l_e^{12}(t)\xi_1\xi_2\right].
\]

\subsubsection{Surface density model in 2D}
\label{section4-1-2}
In this  case,  the function $u(\tbx, t_n)$ in (\ref{Phip}) also satisfies the square-root-Poisson
equation in (\ref{Uepsm23}) which can be truncated on the computational domain ${\mathcal D}$ with homogeneous
Dirichlet boundary conditions as
\bea\label{Fpoisson}
(-\nabla^2)^{1/2}u(\tbx, t_n) = |\phi(\tbx, t_n)|^2,\quad \
\tbx\in{\mathcal D}; \qquad u(\tbx, t_n)|_{\p {\mathcal D}}=0.
\eea
The above problem can be discretized by using a sine pseudospectral method in which the 0-modes
are avoided.  Letting $M,K \in \mathbb{N}$, we denote the index set
\beas
{\mathcal T}_{MK} = \left\{(p, q)\,|\,1\leq p\leq M-1, \ 1\leq q\leq K-1\right\},
\eeas
and define the functions
\[U_{p,q}(\tbx)=\sin(\mu_p^1(\tx-a))\sin(\mu_q^2(\ty-c)),
\qquad (p, q)\in{\mathcal T}_{MK}, \qquad  \tbx=(\tx,\ty)\in {\mathcal D},
\]
where
\bea\label{Eq00}
\mu_p^1 = \fl{p\pi}{b-a}, \qquad \mu_q^2 = \fl{q\pi}{d-c}, \qquad (p, q)\in{\mathcal T}_{MK}.
\eea
Assume that
\bea\label{2D-ansatz0}
u(\tbx, t_n) = \sum_{p=1}^{M-1}\sum_{q=1}^{K-1}
\widehat{u}^s_{pq}(t_n) U_{p,q}(\tbx),  \qquad  \tbx\in {\mathcal D},
\eea
where $\widehat{u}^s_{pq}(t_n)$ is the sine transform of $u(\tbx,t_n)$ at frequencies $(p, q)$.
Substituting (\ref{2D-ansatz0}) into (\ref{Fpoisson}) and taking sine transform on both sides,  we obtain
\bea\label{S-eq0}
\widehat{u}^s_{pq}(t_n) = \fl{(\widehat{|\phi^n|^2})^s_{pq}}{\sqrt{(\mu_p^1)^2+
(\mu_q^2)^2}} , \qquad (p, q) \in {\mathcal T}_{MK},
\eea
where $(\widehat{|\phi^n|^2})^s_{pq}$ is the sine transform of $|\phi(\tbx,t_n)|^2$ at frequencies $(p, q)$.

Combining (\ref{2D-ansatz0}), (\ref{S-eq0}) and (\ref{Phip956}), we obtain
an approximation of $\Phi(\tbx,t)$ in the solution (\ref{2solution}) via
sine spectral method as
\bea\label{phi247}
\qquad \quad
\Phi(\tbx, t) = \sum_{p = 1}^{M-1}\sum_{q = 1}^{K-1}\fl{(\widehat{|\phi^n|^2})^s_{pq}}{\sqrt{(\mu_p^1)^2+
(\mu_q^2)^2}}\left[L(\mu_p^1, \mu_q^2, t) U_{p,q}(\tbx)+l_e^{12}(t)V_{p,q}(\tbx)\right],
\eea
where the functions $L(\xi_1,\xi_2, t)$ and $V_{p,q}(\tbx)$ are  defined as
\beas 
&&L(\xi_1, \xi_2, t)  =  -\left[\left(l_e^{11}(t)-l_e^{33}(t)\right) \xi_1^2+
\left(l_e^{22}(t)-l_e^{33}(t)\right) \xi_2^2\right], \\
&&V_{p,q}(\tbx)=\p_{\tx\ty}U_{p,q}(\tbx)=\mu_p^1\mu_q^2\cos(\mu_p^1(\tx-a))\cos(\mu_q^2(\ty-c)),
\qquad (p, q)\in{\mathcal T}_{MK}.
\eeas
\subsubsection{Approximations in 3D}
\label{section4-2-3}

In 3D case, again the function $u(\tbx, t_n)$ in (\ref{Phip}) also satisfies the Poisson
equation in (\ref{GPPS1789}) which can be truncated on the computational domain ${\mathcal D}$ with homogeneous
Dirichlet boundary conditions as
\bea\label{3D-step1}
-\nabla^2 u(\tbx,t_n) =  |\phi(\tbx, t_n)|^2, \quad\tbx\in{\mathcal D}; \qquad
\qquad u(\tbx, t_n)|_{\p {\mathcal D}}=0.
\eea
The above problem can be discretized by using a sine pseudospectral method in which the 0-modes
are avoided.  Denote the index set
\beas
{\mathcal T}_{MKL} = \left\{(p, q, r)\,|\,1\leq p\leq M-1, \ 1\leq q\leq K-1, \ 1\leq r\leq L-1\right\}
\eeas
where $M, K, L > 0$ are integers and define the functions
\[U_{p,q,r}(\tbx)=\sin(\mu_p^1(\tx-a))\sin(\mu_q^2(\ty-c))\sin(\mu_r^3(\tz-e)),
\quad (p, q,r)\in{\mathcal T}_{MKL},
\]
where
\[\mu_r^3 = r\pi/(f-e), \qquad  1\leq r\leq L-1.\]
Again, we take the (approximate) ansatz
\bea\label{3D-ansatz} \ \
 u(\tbx, t_n) = \sum_{p=1}^{M-1}\sum_{q=1}^{K-1}\sum_{r=1}^{L-1}
\widehat{u}^s_{pqr}(t_n)\; U_{p,q,r}(\tbx), \qquad \tbx=(\tx,\ty,\tz)\in {\mathcal D},
\eea
where $\widehat{u}_{pqr}^s(t_n)$ is the sine transform of $u(\tbx,t_n)$ corresponding to frequencies $(p,
q, r)$.  Substituting (\ref{3D-ansatz}) into the Poisson equation
(\ref{3D-step1}) and noticing the orthogonality of the sine functions,
 we obtain
\bea\label{S-eq1}
\widehat{u}^s_{pqr}(t_n) = \fl{(\widehat{|\phi^n|^2})^s_{pqr}}{(\mu_p^1)^2+
(\mu_q^2)^2 + (\mu_r^3)^2}, \qquad (p, q, r) \in {\mathcal T}_{MKL},
\eea
where $(\widehat{|\phi^n|^2})^s_{pqr}$ is the sine transform of
$|\phi(\tbx,t_n)|^2$ corresponding to frequencies $(p,q,r)$.

Combining (\ref{Phip956}), (\ref{3D-ansatz}) and (\ref{S-eq1}),
we obtain an  approximation of $\Phi(\tbx, t)$ in the solution (\ref{2solution}) via
sine spectral method as
\bea\quad
\Phi(\tbx,t)&=&\sum_{p = 1}^{M-1}\sum_{q = 1}^{K-1}\sum_{r=1}^{L-1} \widehat{u}^s_{pqr}(t_n)
\bigg[L(\mu_p^1, \mu_q^2, \mu_r^3, t)U_{p,q,r}(\tbx)+ l_e^{12}V_{p,q,r}^{(1)}(\tbx) \nonumber\\
&&\qquad +l_e^{13} V_{p,q,r}^{(2)}(\tbx)+l_e^{23}V_{p,q,r}^{(3)}(\tbx) \bigg],
\qquad \tbx\in {\mathcal D},
\eea
where the functions $L(\xi_1, \xi_2, \xi_3, t)$, $V_{p,q,r}^{(1)}(\tbx)$, $V_{p,q,r}^{(2)}(\tbx)$
and   $V_{p,q,r}^{(3)}(\tbx)$ (for $(p, q,r)\in{\mathcal T}_{MKL}$) are defined as
\beas
&&L(\xi_1, \xi_2, \xi_3, t)= -\left[l_e^{11}(t)\xi_1^2 + l_e^{22}(t)\xi_2^2
+l_e^{33}(t)\xi_3^2\right],\\
&&V_{p,q,r}^{(1)}(\tbx)=\p_{\tx\ty}U_{p,q,r}(\tbx)=\mu_p^1\mu_q^2\cos(\mu_p^1(\tx-a))
\cos(\mu_q^2(\ty-c))\sin(\mu_r^3(\tz-e)), \\
&&V_{p,q,r}^{(2)}(\tbx)=\p_{\tx\tz}U_{p,q,r}(\tbx)=\mu_p^1\mu_r^3\cos(\mu_p^1(\tx-a))
\sin(\mu_q^2(\ty-c))\cos(\mu_r^3(\tz-e)), \\
&&V_{p,q,r}^{(3)}(\tbx)=\p_{\ty\tz}U_{p,q,r}(\tbx)=\mu_q^2\mu_r^3\sin(\mu_p^1(\tx-a))
\cos(\mu_q^2(\ty-c))\cos(\mu_r^3(\tz-e)).
\eeas

\begin{remark}\label{Remark2}
After obtaining the numerical solution $\phi(\tbx,t)$ on the bounded computational domain
${\mathcal D}$, if it is needed to recover the original wave function $\psi(\bx,t)$ over
a set of fixed grid points in the Cartesian coordinates $\bx$,
one can  use the standard Fourier/sine interpolation operators from the discrete numerical solution
$\phi(\tbx,t)$ to construct an interpolation continuous function over ${\mathcal D}$ \cite{Boyd1992, Shen},
which can be used to compute $\psi(\bx,t)$ over
a set of fixed grid points in the Cartesian coordinates $\bx$ for any fixed time $t\ge0$.
\end{remark}

\begin{remark}\label{Remark3}
If the potential $V(\bx)$ in (\ref{GGPE}) is replaced by a time-dependent potential, e.g.
$V(\bx,t)$, the rotating Lagrangian coordinates transformation and the numerical method
are still valid provided that we replace $W(\tbx,t)$ in (\ref{wtbx534}) by
$W(\tbx,t) = V(A(t)\tbx,t)$ for $\tbx\in {\mathbb R}^d$ and $t\ge0$.
\end{remark}

\section{Numerical results}
\setcounter{equation}{0}
\label{section5}

In this section, we first test the accuracy of our numerical method, where throughout we apply the two-dimensional surface
density model. Then study the dynamics of  rotating dipolar BECs, including the center of mass, angular momentum expectation and condensate widths. In addition, the dynamics of vortex lattices in rotating dipolar BEC are presented.

\subsection{Numerical accuracy}

In order to test numerical accuracy,  we consider a 2D GPE (\ref{CGPE})-(\ref{Cvarphi})
with the SDM long-range interaction  (\ref{Uepsm23}) and harmonic potential (\ref{Vpoten}),
i.e.\ $d = 2$ in the GPE (\ref{CGPE}). The other
parameters are chosen as   $\Omega = 0.4$, $\gamma_x = \gamma_y = 1$, $\eta = -\fl{15}{2}$
and dipole axis  ${\bn} = (0, 0, 1)^T$.   The
initial condition  in (\ref{Cinitial}) is taken as
\bea
\phi_0(\tbx) = \fl{1}{\pi^{1/4}} e^{\fl{-({\tilde x}^2 + 2{\tilde y}^2)}{2}}, \qquad \tbx\in{\mathcal D},
\eea
where we perform our simulations on the bounded computational domain ${\mathcal D} = [-16, 16]^2$.
Denote $\phi^{(\Dt\tx, \Dt\ty, \Dt t)}(t)$ as the numerical solution at time $t$ obtained with the mesh size
$(\Dt\tx, \Dt\ty)$ and time step $\Dt t$.
With a slight abuse of notation, we let $\phi(t)$ represent the numerical solution with very fine mesh size $\Dt\tx =
\Dt\ty = 1/64$ and small time step  $\Dt t =0.0001$ and assume it to be a
sufficiently good representation of the {\it
exact} solution at time $t$.

\begin{table}
\begin{center}
\begin{tabular}{|l|c|c|c|c|c|c|c|}
\hline
 & $\Delta \widetilde{x} = 1/2$ & $\Delta \widetilde{x} = 1/4$ & $\Delta \widetilde{x} = 1/8$ & $\Delta \widetilde{x} =
1/16$ \\
  \hline
$\bt = 33.5914$ &6.1569E-2& 1.7525E-4& 5.8652E-11 &$<$1E-11 \\
 \hline
$\bt = 58.7849$ &1.9746E-1&2.3333E-3&2.5738E-8&2.6124E-11\\
 \hline
$\bt = 92.3762$ &4.8133E-1&1.3385E-2&1.6620E-6&6.2264E-10\\
\hline
$\bt = 119.8488$ &1.2984&7.7206E-2&9.5202E-5&3.0974E-8\\
 \hline
\end{tabular}\\
\caption{Spatial discretization errors $\|\phi(t)-\phi^{(\Dt \tx, \Dt \ty, \Dt t)}(t)\|$ at time $t  = 1$.
}\label{T1}
\end{center}
\end{table}

\begin{table}
\begin{center}
\begin{tabular}{|l|c|c|c|c|c|c|}
\hline
 & $\Delta t = 1/40$ & $\Delta t = 1/80$ & $\Delta t = 1/160$ & $\Delta t = 1/320$ &
 $\Delta t = 1/640$  \\
 \hline
$\bt = 33.5914$ &1.0434E-3&2.6018E-4&6.4992E-5&1.6233E-5&4.0456E-6\\
 \hline
$\bt = 58.7849$ &2.5241E-3&6.2783E-4&1.5674E-4&3.9143E-5&9.7550E-6\\
 \hline
$\bt = 92.3762$ &4.9982E-3&1.2380E-3&3.0882E-4&7.7108E-5&1.9215E-5\\
 \hline
 $\bt = 119.8488$ &1.1417E-2&2.7716E-3&6.9009E-4&1.7223E-4&4.2915E-5\\
 \hline
\end{tabular}\\
\caption{Temporal discretization errors $\|\phi(t)-\phi^{(\Dt \tx, \Dt \ty, \Dt t)}(t)\|$ at time $t  =
1$.}\label{T2}
\end{center}
\end{table}

Tables \ref{T1}--\ref{T2} show the spatial and temporal errors of our numerical method for different  $\bt$
in the GPE (\ref{CGPE}), where the errors are computed as $\|\phi(t) - \phi^{(\Dt\tx, \Dt\ty, \Dt t)}(t)\|_{l^2}$
(with $\Delta \widetilde{x} = \Delta \widetilde{y}$) at time  $t = 1$. To calculate the spatial errors in Table
\ref{T1}, we always use  a very small time step $\Dt  t = 0.0001$ so that the errors from time discretization can be
neglected compared to those from spatial discretization. Table \ref{T1} shows that the spatial accuracy of
our method  is of spectral order. In addition, the spatial errors increase with the nonlinearity
coefficient $\bt$ when the mesh size is kept constant.

In Table \ref{T2}, we always use mesh sizes $\Dt\tx = \Dt\ty = 1/64$ which are the same as
those used in obtaining the `exact' solution, so that one can regard the spatial discretization as `exact'
and the only errors are from time discretization. For different $\bt$, Table \ref{T2} shows second order
decrease of the temporal errors with respect to time-step size $\Dt t$. Similarly, for the same $\Dt t$,
the temporal errors increase with $\bt$.


\subsection{Dynamics of center of mass}

In the following, we study the dynamics of the center of mass   by
directly simulating the GPE (\ref{GGPE})--(\ref{GGPE93}) in 2D with SDM long-range interaction (\ref{Uepsm23})
and harmonic potential (\ref{Vpoten}).
To that end, we take $d = 2$, $\bt =
30\sqrt{{10}/{\pi}}$, $\eta = -\fl{15}{2}$ and  dipole axis
$\bn = (1, 0, 0)^T$.  The initial condition in (\ref{Ginitial}) is taken as
\be\label{example-initial}
\phi_0(\bx) = \ap\,\zeta(\bx - \bx^0), \qquad\mbox{with}\quad\zeta(\bx) = (x + iy)
e^{\fl{-(x^2 + y^2)}{2}}, \quad
\bx\in {\mathcal D},
\ee
where the constant $\ap$ is chosen to satisfy the normalization condition $\|\psi_0\|^2 = 1$.
Initially, we take $\bx^0 = (1, 1)^T$.
In our simulations, we use the computational domain ${\mathcal D} = [-16,16]^2$,
 the mesh size  $\Dt\tx = \Dt\ty = 1/16$ and the time step size $\Dt  t= 0.0001$.

\begin{figure}[h!]
\centerline{
\psfig{figure=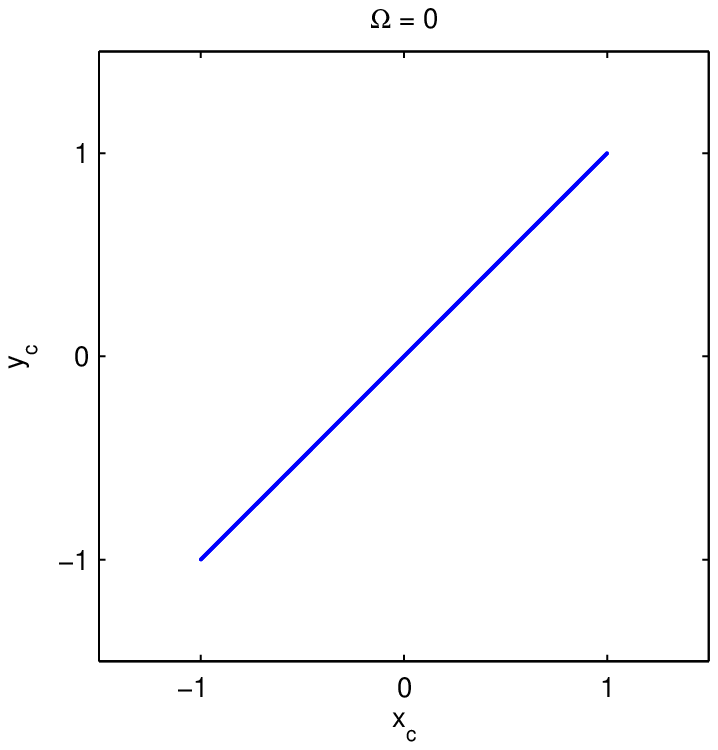,height=4.266cm,width=4.3cm,angle=0}\quad \
\psfig{figure=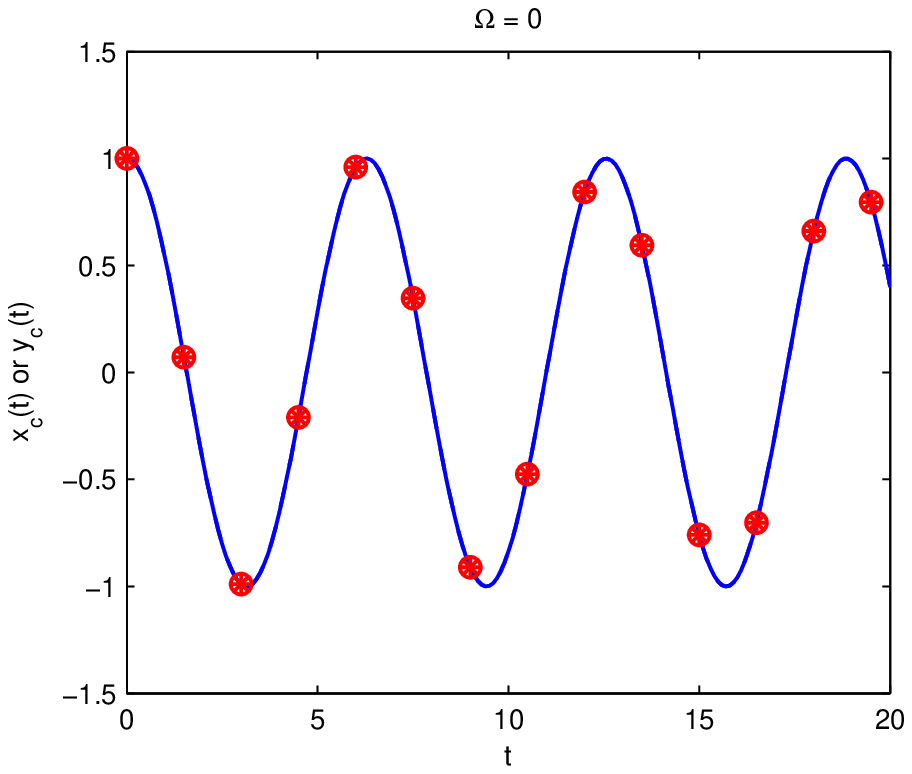,height=4.2cm,width=6.3cm,angle=0}
}
\centerline{
\psfig{figure=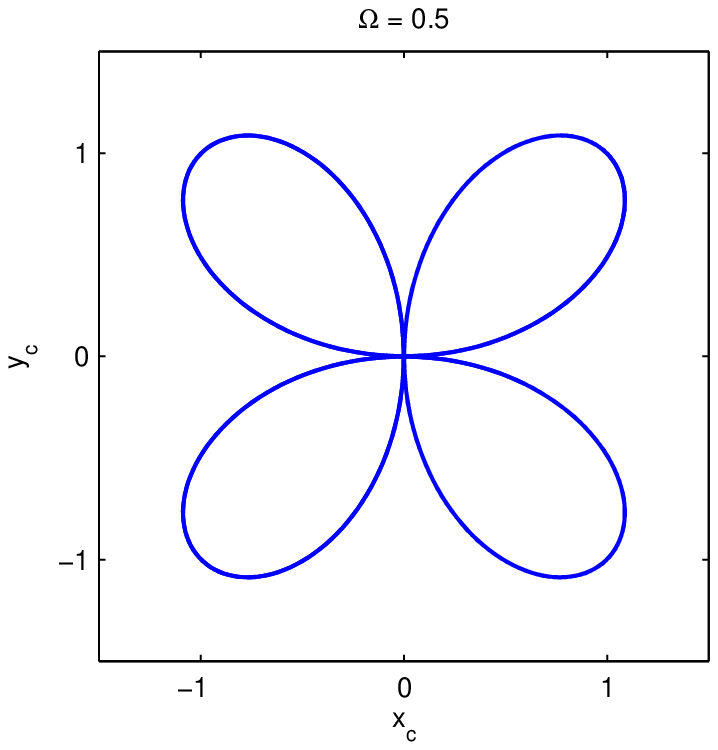,height=4.266cm,width=4.3cm,angle=0}\quad \
\psfig{figure=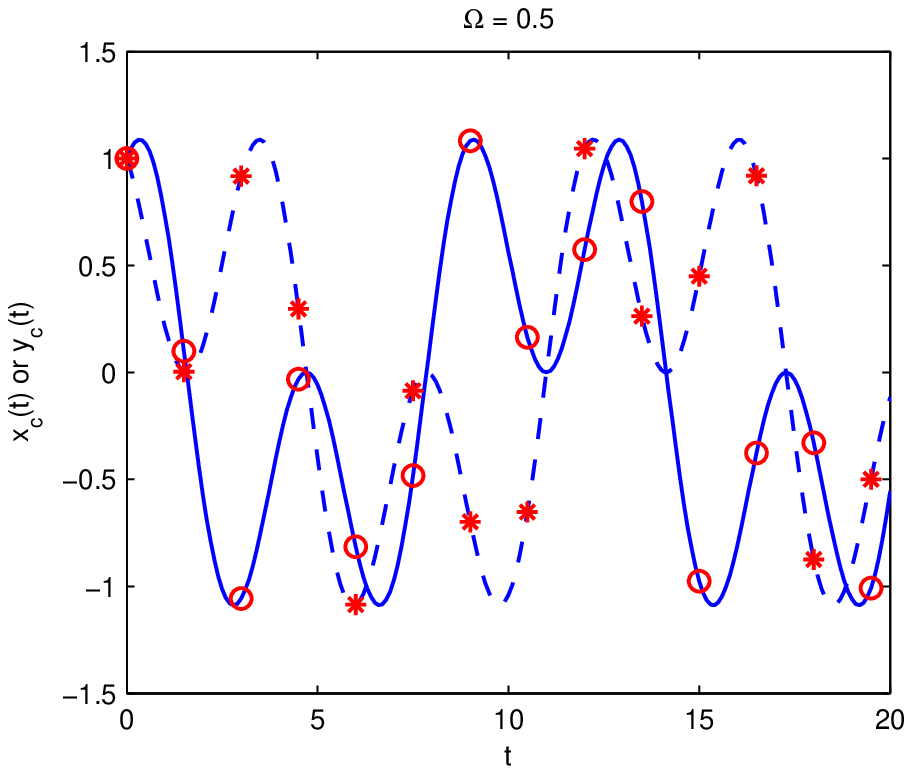,height=4.2cm,width=6.3cm,angle=0}
}
\centerline{
\psfig{figure=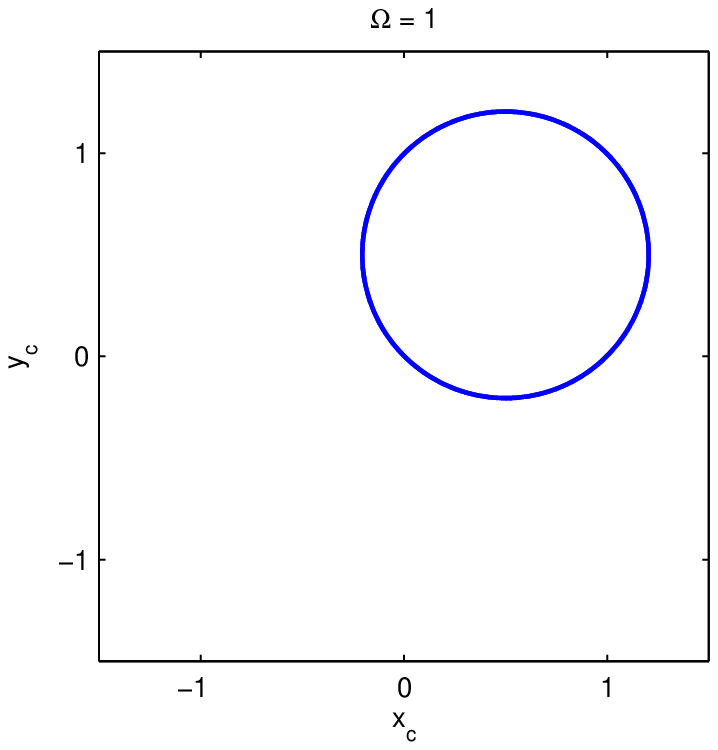,height=4.266cm,width=4.3cm,angle=0}\quad \
\psfig{figure=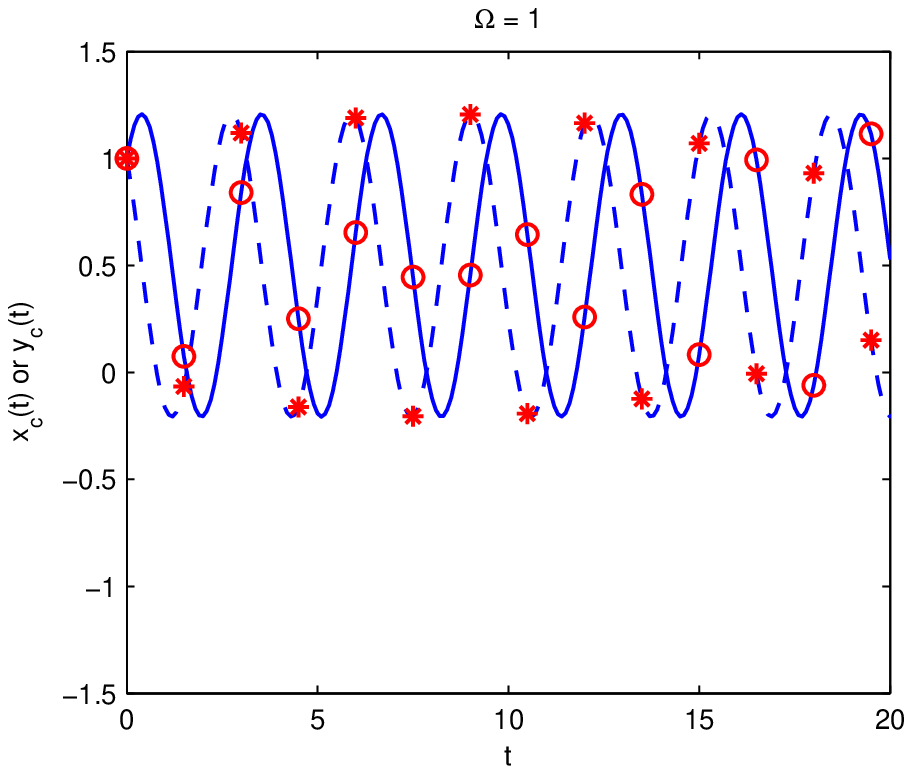,height=4.2cm,width=6.3cm,angle=0}
}
\centerline{
\psfig{figure=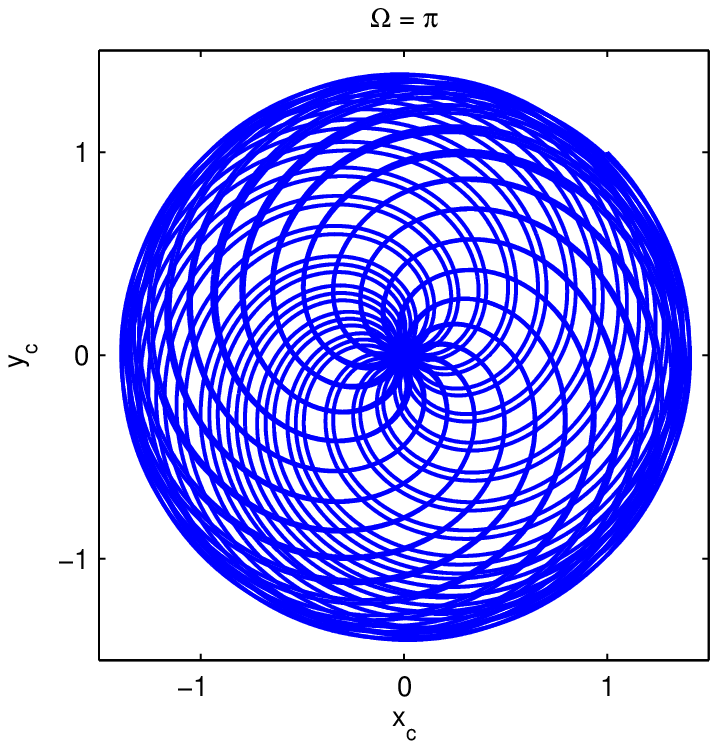,height=4.266cm,width=4.3cm,angle=0}\quad \
\psfig{figure=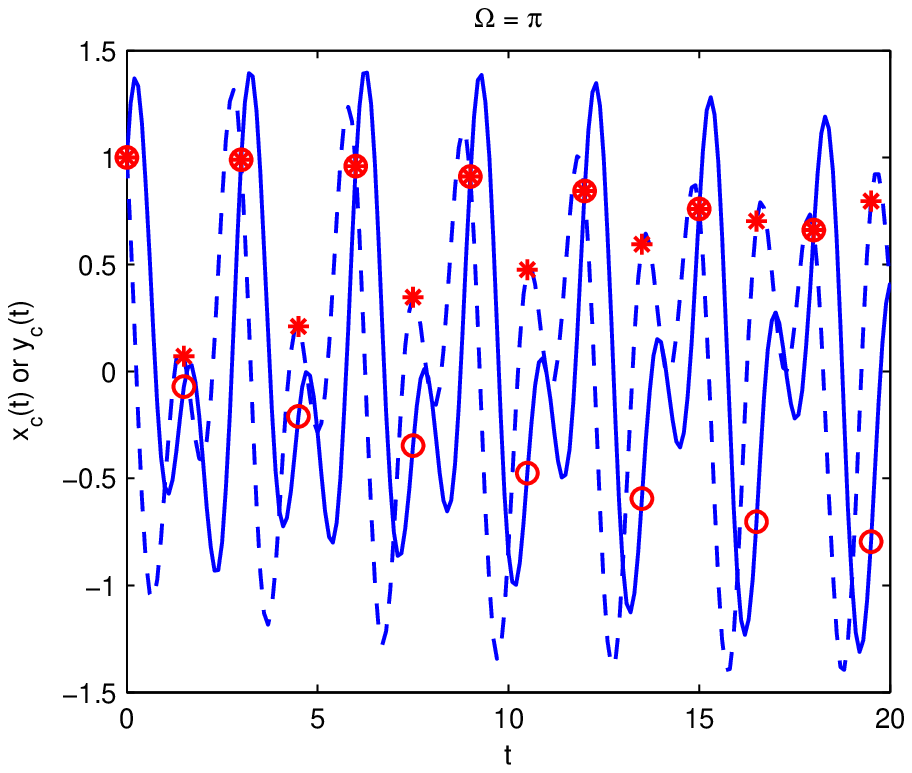,height=4.2cm,width=6.3cm,angle=0}
}
\caption{Results for $\gamma_x = \gamma_y = 1$.
Left:   trajectory of the center of mass,  $\bx_c(t) = (x_c(t), y_c(t))^T$ for $0\leq t \leq 100$.
Right:  coordinates of the trajectory $\bx_c(t)$ (solid line: $x_c(t)$,
dashed line: $y_c(t)$) for different rotation speed $\Omega$, where the solid and dashed lines
are obtained by directly simulating
the GPE and `*' and `o' represent the solutions to
the ODEs in Lemma \ref{lemma4}.}\label{F1}
\end{figure}

We consider the following two sets of trapping frequencies:  (i) $\gamma_x = \gamma_y = 1$, and
(ii) $\gamma_x = 1$, $\gamma_y = 1.1$.  Figure \ref{F1} shows the trajectory of the center of mass
$\bx_c(t)$ in the original coordinates as well as
the time evolution of its coordinates for different angular velocities
$\Omega$, where $\gm_x = \gm_y = 1$.  On the other hand, Figure \ref{F2} presents the same quantities for $\gm_x = 1$
and $\gm_y = 1.1$.
In addition,  the numerical results are  compared with analytical ones from solving the ODEs in
(\ref{ODE2})--(\ref{ODE1}). Figs. \ref{F1}--\ref{F2} show that if the
external trap is symmetric, i.e.\ $\gm_x = \gm_y$, the center of mass always moves
within a bounded region which is symmetric with respect to the trap center $(0,0)^T$. Furthermore, if the angular
velocity $\Og$ is rational, the movement is periodic with a period depending on both the angular velocity and the
trapping frequencies.
In contrast, when $\gm_x\neq\gm_y$, the dynamics of the center of mass become more complicated.
The simulation results in Figs. \ref{F1}--\ref{F2} are consistent with those obtained by solving the ODE system
in Lemma \ref{lemma4} for given $\Og$, $\gamma_x$, and $\gamma_y$ \cite{Zhang2007} and those numerical results
reported in the literatures by other numerical methods \cite{Bao2006,Bao2006-1,Bao2009}.

On the other hand, we also study the dynamics of the center of mass $\tbx_c(t)$ in the new coordinates.
When $\gm_x =
\gm_y$ and $\Og$ arbitrary,  the center of mass has a periodic motion on the straight line segment connecting
$-\tbx^0$ and $\tbx^0$. This is also true for $\bx_c(t)$ with $\Og =0$ (cf.\ Fig. \ref{F1}). However, the trajectories
are different for different  $\Og$  if $\gm_x \neq \gm_y$.  This observations agree with the results in
Lemma \ref{lemma5}.

In addition, our simulations  show  that the dynamics of the center of mass are independent of the interaction
coefficients $\bt$ and $\eta$, which is consistent with Lemma \ref{lemma4}.

\begin{figure}[h!]
\centerline{
\psfig{figure=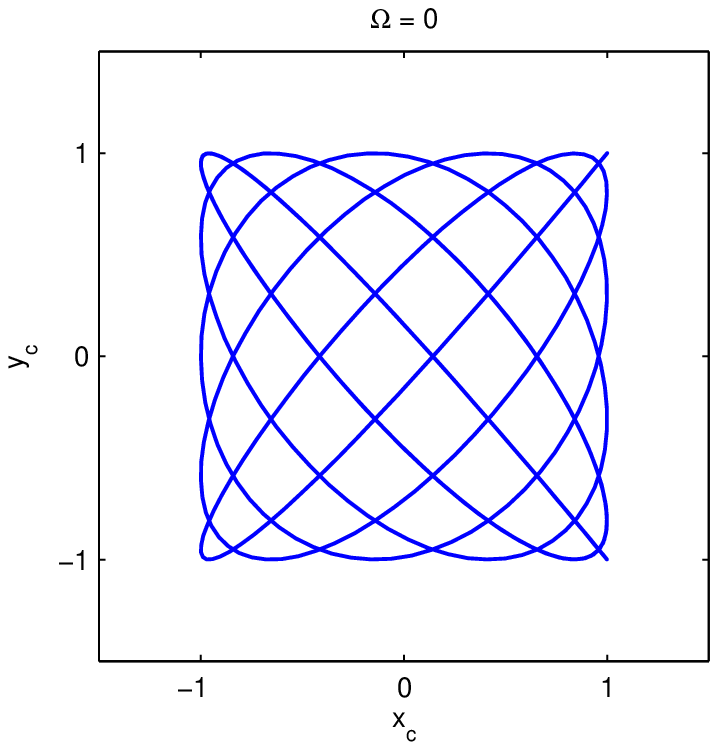,height=4.266cm,width=4.3cm,angle=0}\quad \
\psfig{figure=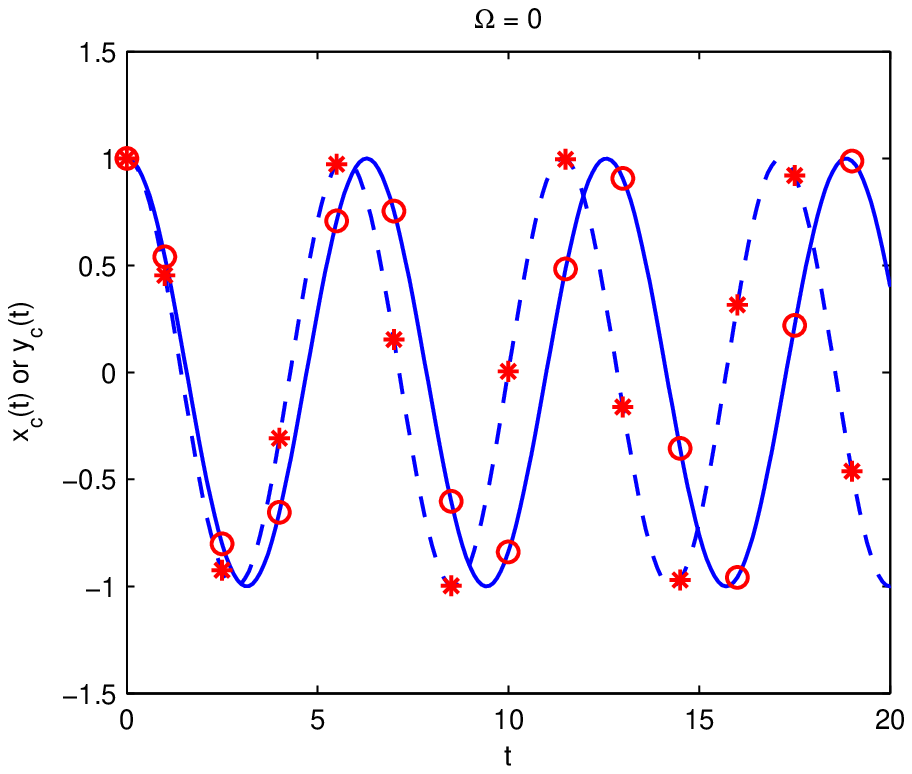,height=4.2cm,width=6.3cm,angle=0}
}
\centerline{
\psfig{figure=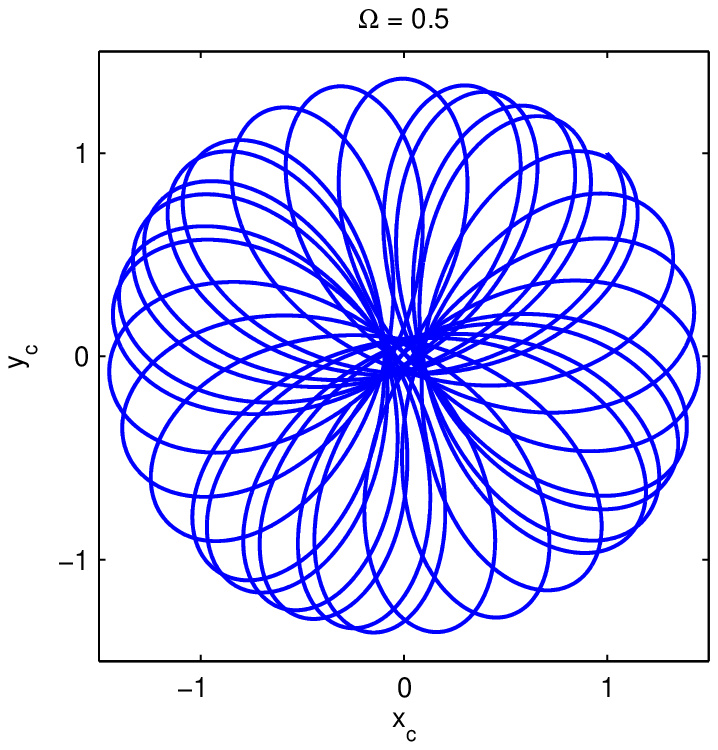,height=4.266cm,width=4.3cm,angle=0}\quad \
\psfig{figure=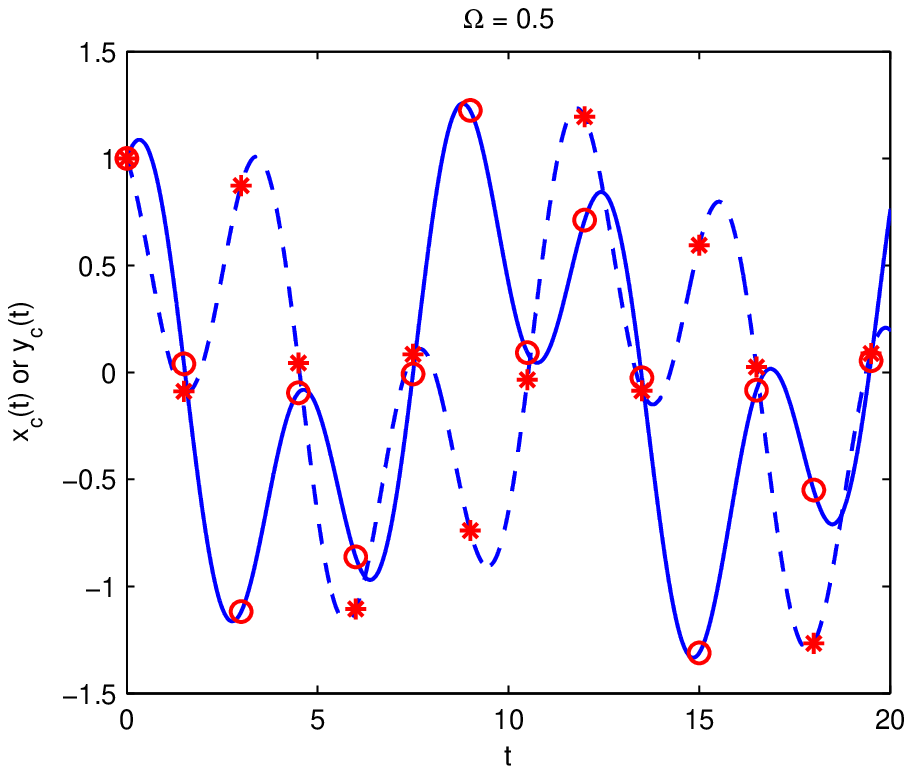,height=4.2cm,width=6.3cm,angle=0}
}
\centerline{
\psfig{figure=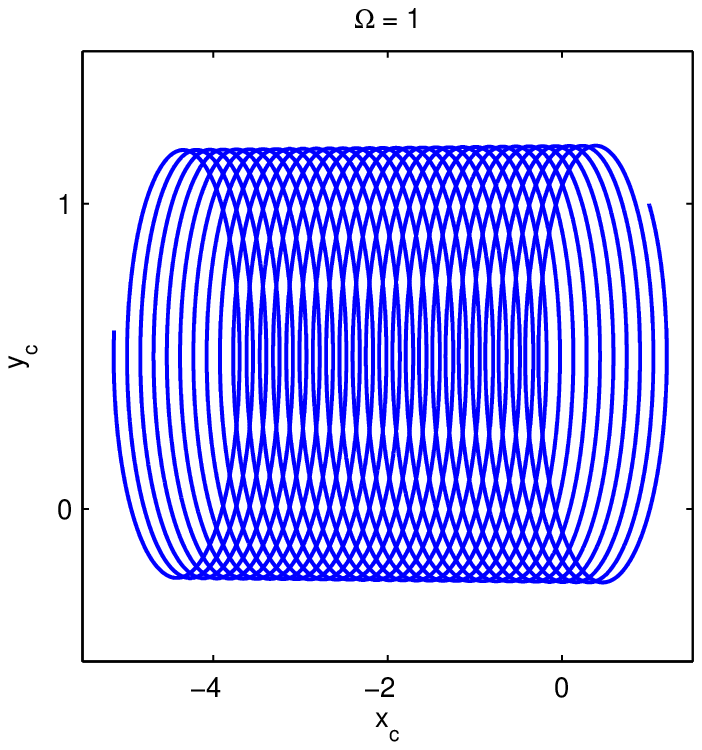,height=4.266cm,width=4.3cm,angle=0}\quad \
\psfig{figure=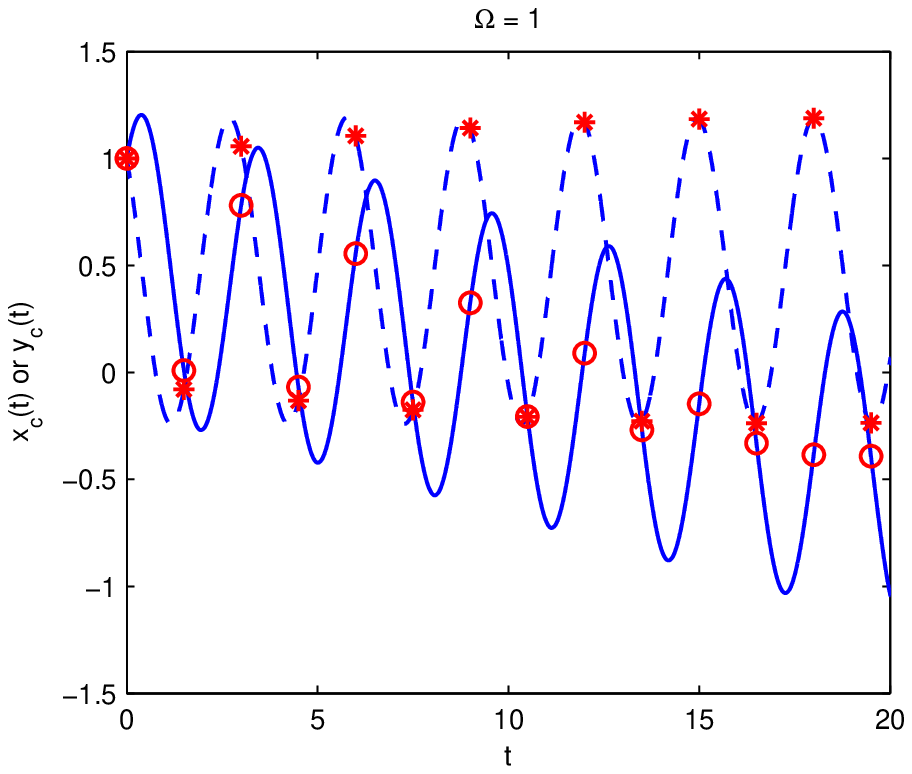,height=4.2cm,width=6.3cm,angle=0}
}
\centerline{
\psfig{figure=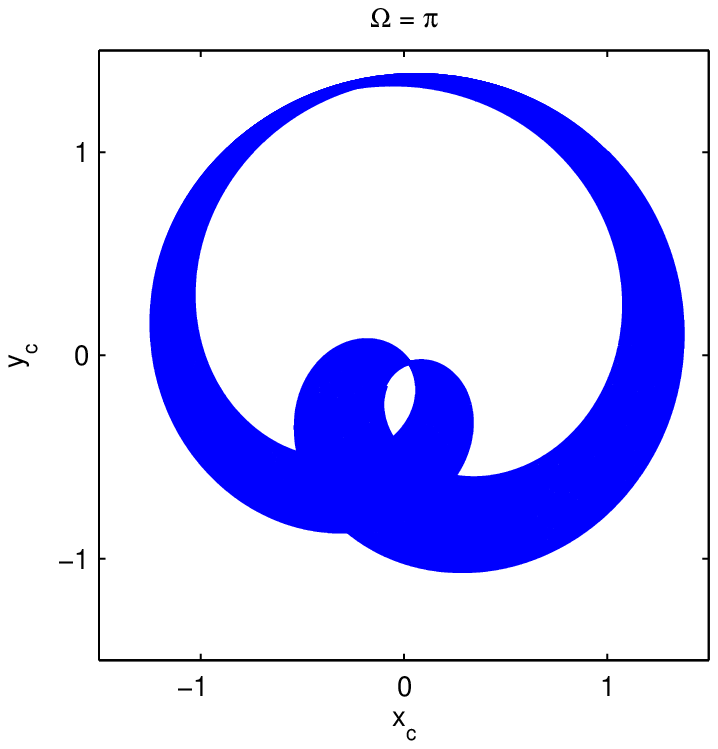,height=4.266cm,width=4.3cm,angle=0}\quad \
\psfig{figure=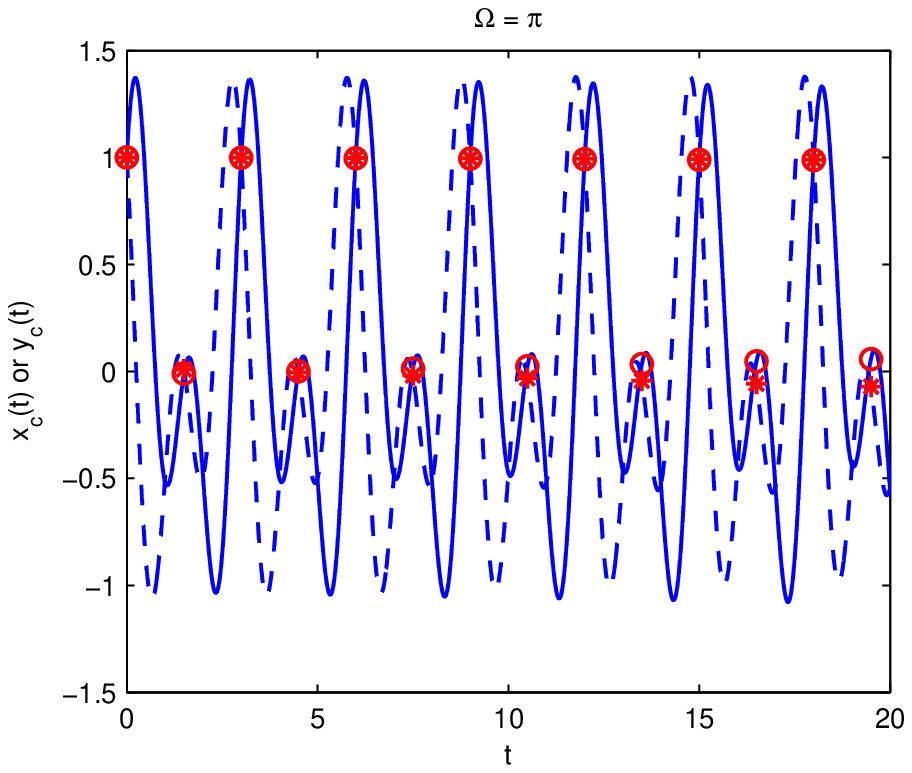,height=4.2cm,width=6.3cm,angle=0}
}
\caption{Results for $\gamma_x = 1, \gamma_y = 1.1$.
Left:   trajectory of the center of mass,  $\bx_c(t) = (x_c(t), y_c(t))^T$ for $0\leq t \leq 100$.
Right:  coordinates of the trajectory $\bx_c(t)$ (solid line: $x_c(t)$,
dashed line: $y_c(t)$) for different rotation speed $\Omega$, where the solid and dashed lines
are obtained by directly simulating
the GPE and `*' and `o' represent the solutions to
the ODEs in Lemma \ref{lemma4}.}
\end{figure}\label{F2}

\subsection{Dynamics of angular momentum expectation and condensate widths}

To study the dynamics of the angular momentum expectation and condensate widths,
we adapt the GPE (\ref{GGPE})--(\ref{GGPE93}) in 2D with SDM long-range interaction (\ref{Uepsm23})
and harmonic potential (\ref{Vpoten}), i.e.
we take  $d = 2$ and $\Og = 0.7$.
Similarly, the initial condition in (\ref{Ginitial}) is taken as
\be
\psi_0(\bx) = \ap\,\zeta(\bx), \qquad \bx \in{\mathcal D},
\ee
where $\zeta(\bx)$ is defined in (\ref{example-initial}) and $\ap$ is a constant such that
$\|\psi_0\|^2 = 1$.  In our simulations, we consider the following four cases:
\begin{enumerate}
\item[(i) ] $\gm_x = \gm_y = 1$, $\beta = 25\sqrt{{10}/{\pi}}$, $\eta = 0$, and $\bn = (1, 0, 0)^T$;
\item[(ii) ] $\gm_x = \gm_y = 1$, $\beta = 25\sqrt{{10}/{\pi}}$, $\eta = -15$,  and $\bn = (1, 0, 0)^T$;
\item[(iii) ] $\gm_x = \gm_y = 1$, $\beta = 55\sqrt{{10}/{\pi}}$, $\eta = -15$,  and $\bn = (0, 0, 1)^T$;
\item[(iv) ] $\gm_x = 1$,  $\gm_y = 1.1$, $\beta = 55\sqrt{{10}/{\pi}}$, $\eta = -15$,  and $\bn = (0, 0, 1)^T$.
\end{enumerate}

\begin{figure}[h!]
\centerline{
\psfig{figure=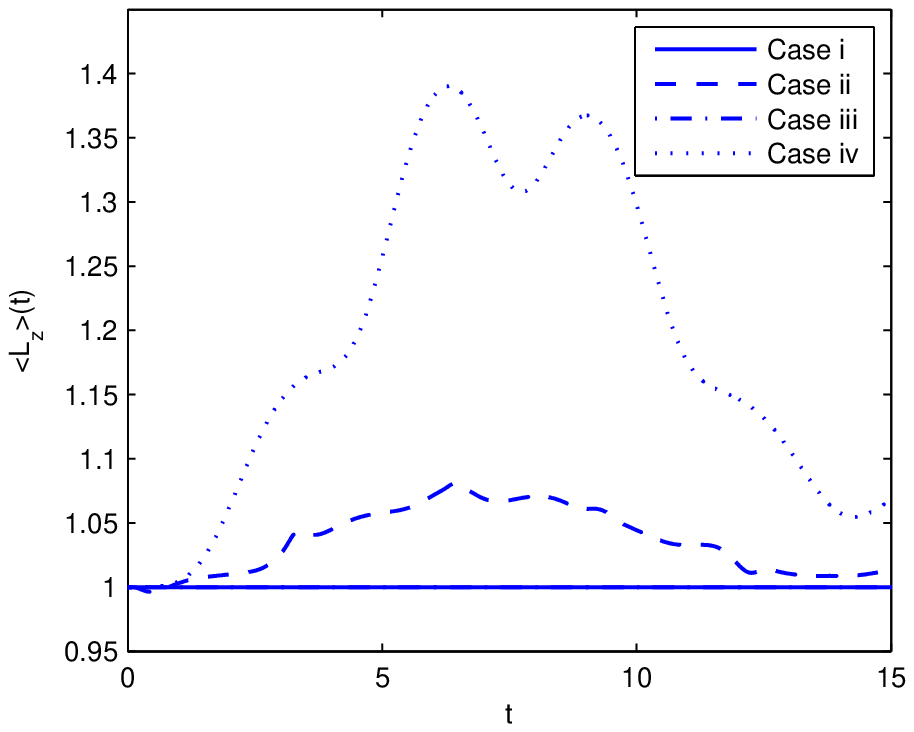,height=5.5cm,width=6.5cm,angle=0}\quad
\psfig{figure=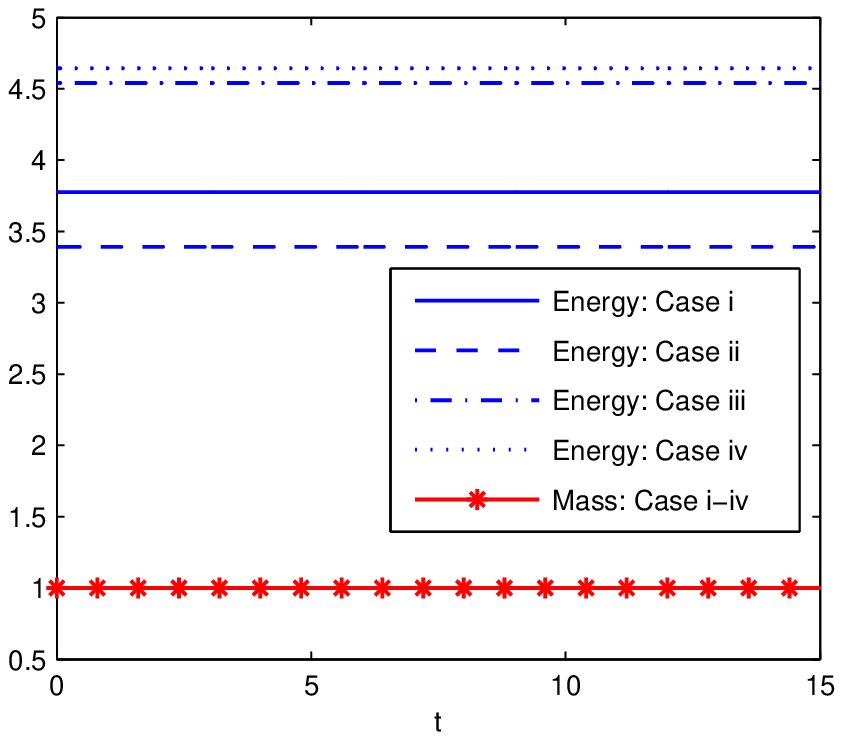,height=5.5cm,width=6.5cm,angle=0}}
\caption{Time evolution of the angular momentum expectation (left) and energy and mass (right)
for Cases (i)-(iv) in section 5.3.}
\label{F3}
\end{figure}

\begin{figure}[h!]
\centerline{
i) \psfig{figure=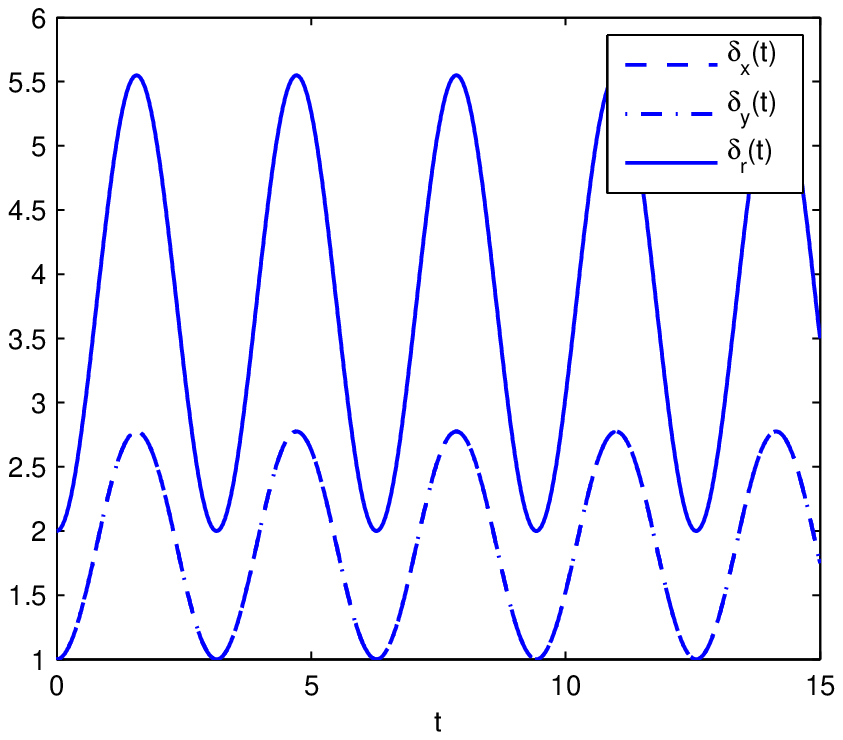,height=4.5cm,width=6.2cm,angle=0}
ii) \psfig{figure=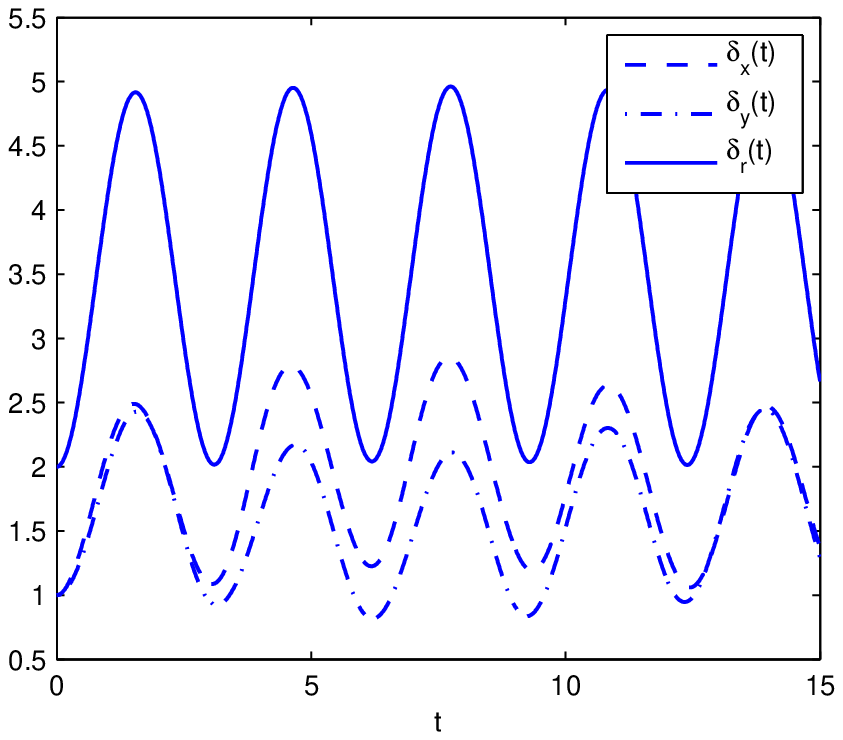,height=4.5cm,width=6.2cm,angle=0}
}
\centerline{
iii) \psfig{figure=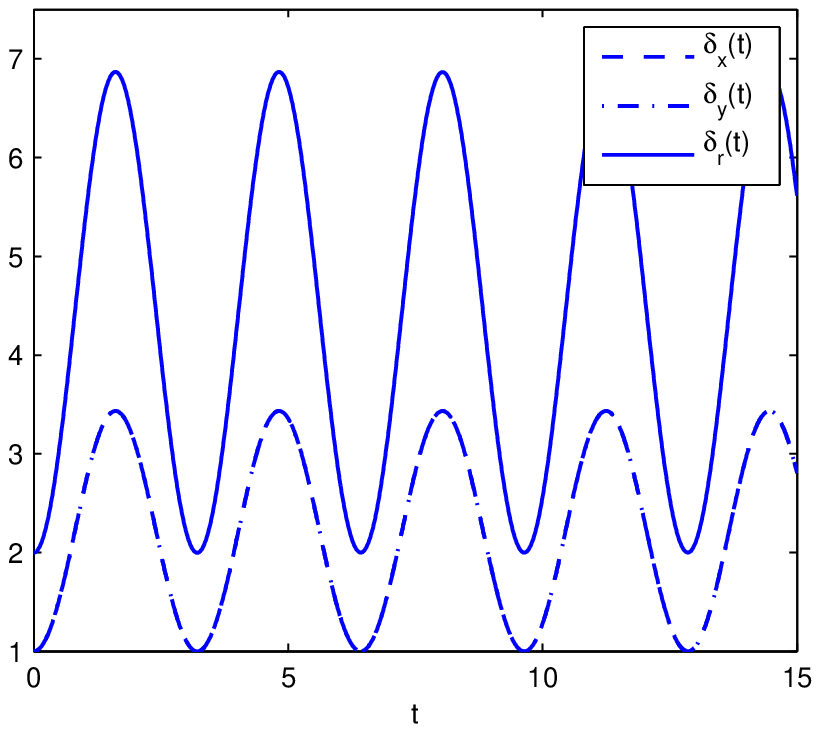,height=4.5cm,width=6.2cm,angle=0}
iv)\psfig{figure=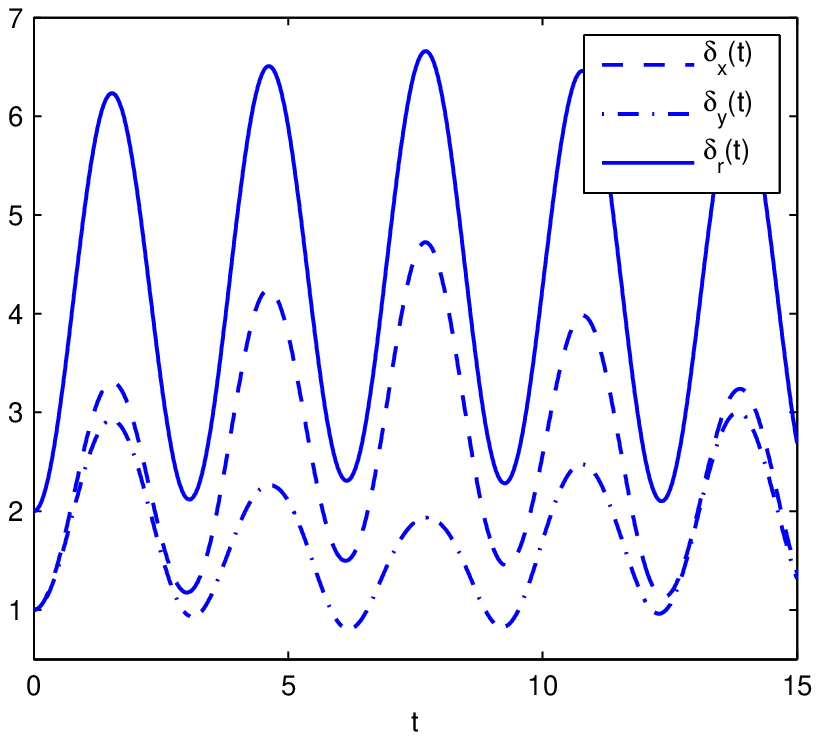,height=4.5cm,width=6.2cm,angle=0}
}
\caption{Time evolution of  condensate widths in the Cases (i)--(iv) in section 5.3.}\label{F4}
\end{figure}

In Figure \ref{F3},  we present the dynamics of the angular momentum expectation,
energy and mass  for each of the above four cases in the interval $t\in[0, 15]$.  We see that
if  the external trap is radially symmetric in 2D, then the angular momentum expectation is
conserved when either there is no dipolar interaction (Case (i)) or  the dipolar axis
is parallel to the $z$-axis (Case (iii)).  Otherwise, the angular momentum expectation is
not conserved.  The above numerical observations are consistent
with the analytical results obtained in Lemma \ref{lemma1}.  In addition,  we find that
our method conserves the energy and mass very well during the dynamics
(cf. Fig.\ \ref{F3} right). Furthermore, from our additional numerical results
not shown here for brevity, we observed that the angular momentum expectation
is conserved in 3D for any initial data if  the external trap is cylindrically
symmetric and either there is no dipolar interaction or  the dipolar axis
is parallel to the $z$-axis, which can also be justified mathematically.

The dynamics of the condensate widths are presented in Figure \ref{F4}. We find
that  $\delta_r(t)$  is periodic as long as the trapping frequencies satisfy
$\gamma_x = \gamma_y$ and the influence of the dipole axis vanishes, e.g.\ in the Case (i),
which confirms the analytical results of Lemma \ref{lemma2}.
Furthermore, from our additional numerical results not shown here for brevity,
we observed that $\dt_r(t)$ is periodic and  $\dt_x(t) = \dt_y(t) = \fl{1}{2}\dt_r(t)$
if $\eta = 0$ for any initial data or ${\bf n} = (0, 0, 1)^T$
for radially symmetric or central vortex-type initial data.

\begin{figure}[htb!]
\centerline{
\psfig{figure=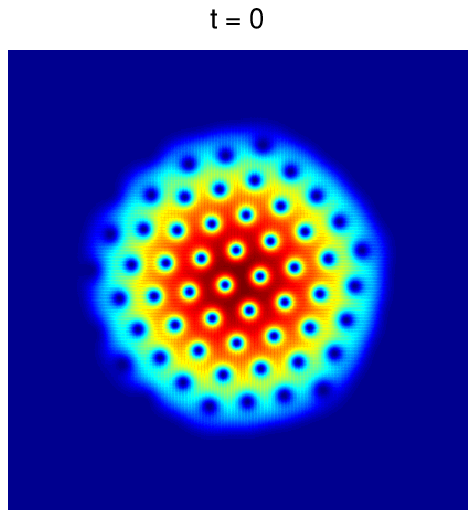,height=4.5cm,width=4.5cm,angle=0}
\psfig{figure=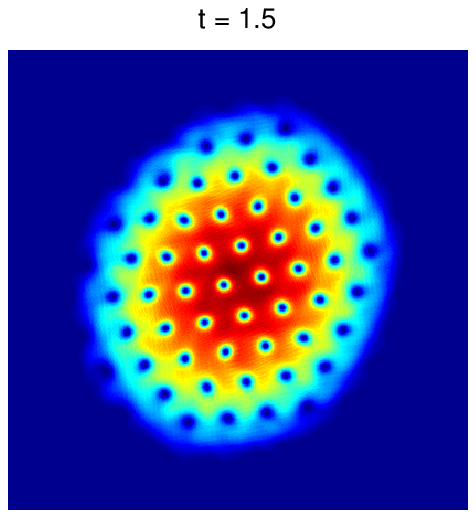,height=4.5cm,width=4.5cm,angle=0}
\psfig{figure=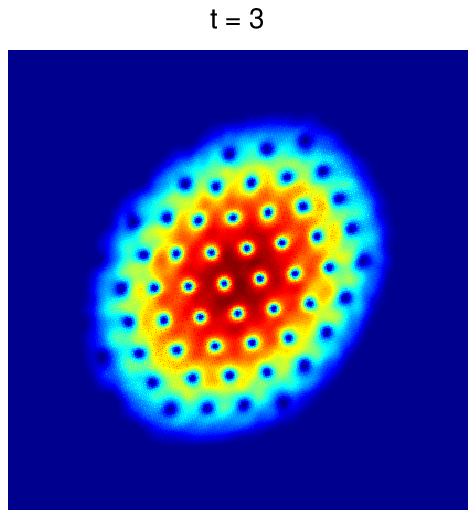,height=4.5cm,width=4.5cm,angle=0}
}
\centerline{
\psfig{figure=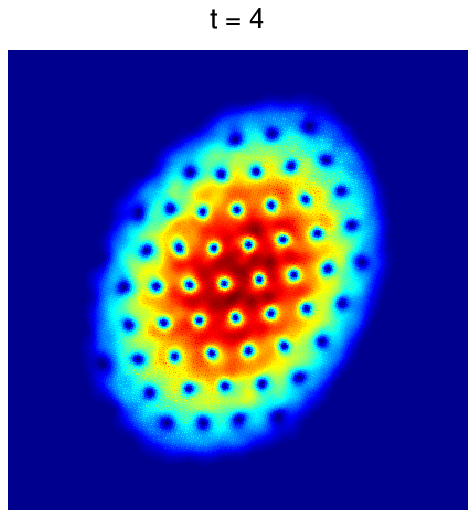,height=4.5cm,width=4.5cm,angle=0}
\psfig{figure=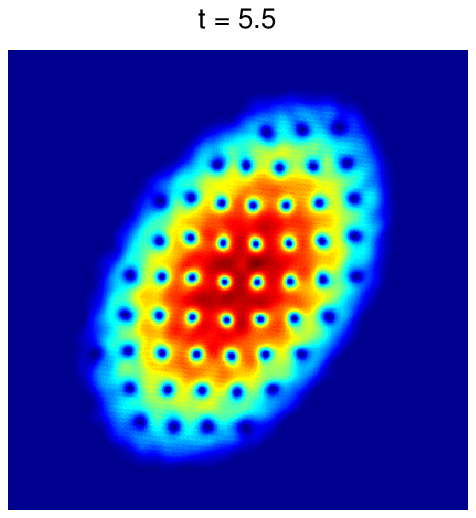,height=4.5cm,width=4.5cm,angle=0}
\psfig{figure=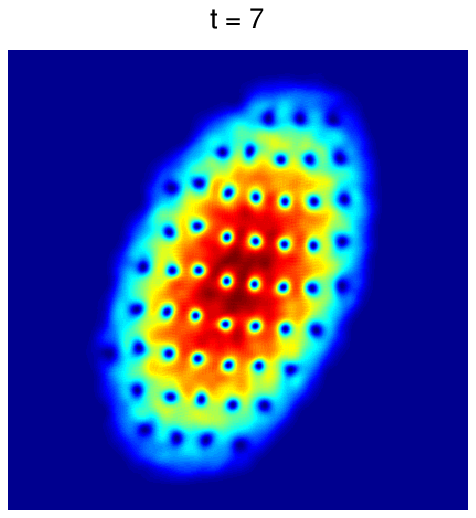,height=4.5cm,width=4.5cm,angle=0}
}
\caption{Contour plots of the density function $|\psi(\bx,t)|^2$ for dynamics of a vortex lattice in a rotating
BEC (Case (i)). Domain displayed:  $(x,y)\in[-13,13]^2$.}\label{F5-0}
\end{figure}

\subsection{Dynamics of quantized vortex lattices}

\begin{figure}[htb!]
\centerline{
\psfig{figure=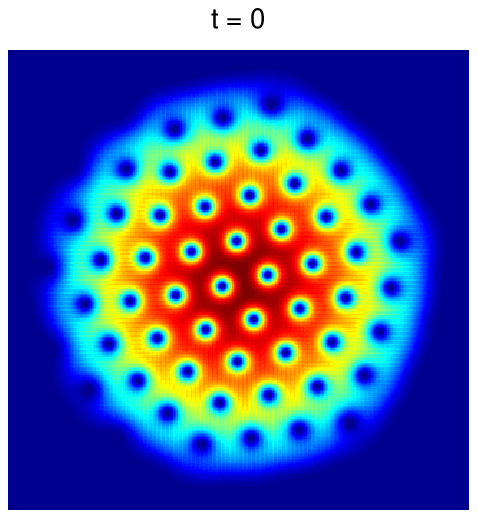,height=4.5cm,width=4.5cm,angle=0}
\psfig{figure=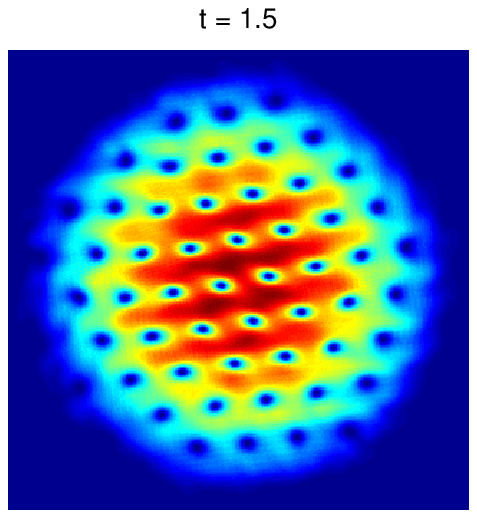,height=4.5cm,width=4.5cm,angle=0}
\psfig{figure=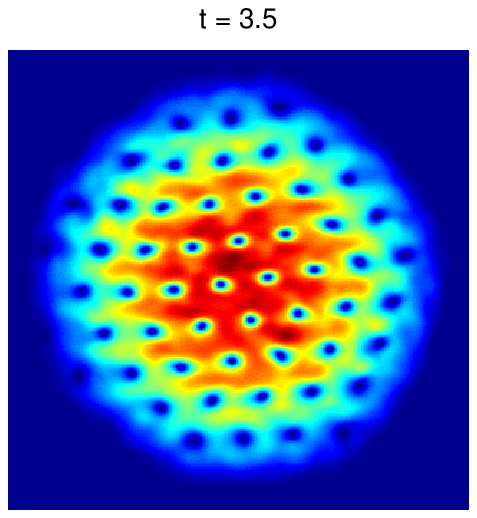,height=4.5cm,width=4.5cm,angle=0}
}
\centerline{
\psfig{figure=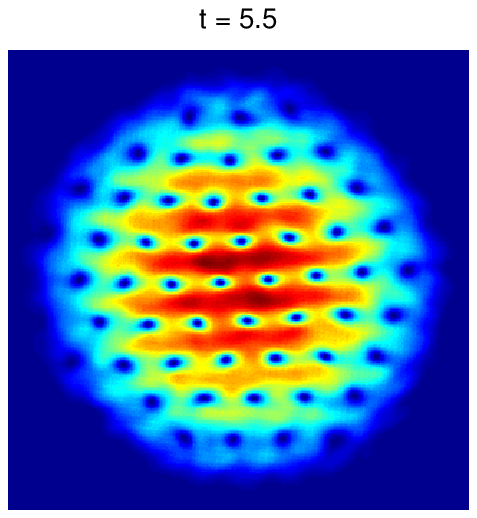,height=4.5cm,width=4.5cm,angle=0}
\psfig{figure=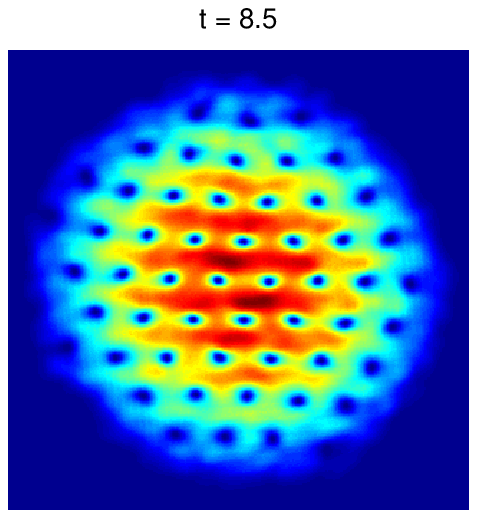,height=4.5cm,width=4.5cm,angle=0}
\psfig{figure=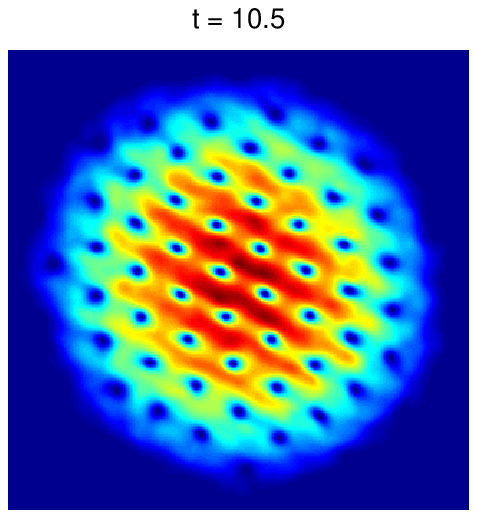,height=4.5cm,width=4.5cm,angle=0}
}
\caption{Contour plots of the density function $|\psi(\bx,t)|^2$ for dynamics of a vortex lattice in a rotating
dipolar BEC (Case (ii)). Domain displayed:  $(x,y)\in[-10,10]^2$.}\label{F5}
\end{figure}

In the following, we apply our numerical method to study the dynamics of quantized vortex lattices
in rotating dipolar BECs. Again, we adapt the GPE (\ref{GGPE})--(\ref{GGPE93}) in
2D with SDM long-range interaction  (\ref{Uepsm23}) and harmonic potential (\ref{Vpoten}), i.e.
we choose $d = 2$, $\bt = 1000$ and $\Og = 0.9$.
The initial datum in (\ref{Ginitial}) is chosen as a
stationary vortex lattice which is computed numerically by using
the method in \cite{Zeng2009, Zhang2010} with the above parameters and
$\gm_x = \gm_y = 1$, $\eta = 0$, i.e. no long-range dipole-dipole interaction initially.
Then the dynamics of vortex lattices are studied in  two cases:
\begin{enumerate}
\item[(i)]  perturb  the external potential by setting $\gm_x = 1.05$ and $\gm_y = 0.95$ at $t=0$;
\item[(ii)] turn on the dipolar interactions by setting $\eta = -600$ and dipolar axis
$\bn = (1, 0, 0)^T$ at time $t=0$.
\end{enumerate}
In our simulations, we use ${\mathcal D} = [-16, 16]^2$, $\Dt\tx = \Dt\ty = 1/16$ and $\Dt t =
0.0001$. Figures \ref{F5-0}--\ref{F5}  show the contour plots of
the density function $|\psi(\bx,t)|^2$ at different time
steps for Cases (i) and (ii), respectively, where the wave function $\psi(\bx,t)$  is obtained from
$\phi(\tbx,t)$ by using interpolation via sine basis
(see Remark \ref{Remark2}).  We see that during the  dynamics, the number
of vortices is conserved in both cases. The lattices rotate to form different patterns because of the
anisotropic external potential and dipolar interaction in Cases (i) and (ii), respectively.
In addition, the results
in Case (i) are similar to those obtained in \cite{Bao2006}, where a spectral type method in polar
coordinates was used to simulate the dynamics  of vortex lattices.

\vskip 20pt

\section{Conclusions}
\label{section6}

We proposed a simple and efficient numerical method to simulate the dynamics of rotating dipolar
Bose-Einstein condensation (BEC) whose properties are  described by the Gross--Pitaevskii equation (GPE)
with both the angular rotation term and the long-range dipole-dipole
interaction.  First, by decoupling the short-range and long-range interactions, we
reformulate the GPE as a Gross-Pitaevskii-(fractional) Poisson system. Then we
eliminate the angular rotation term from the GPE using a rotating Lagrangian
coordinate transformation,  which makes it possible to design a
simple and efficient numerical method. In the new rotating Lagrangian coordinates, we presented  a
numerical method which combines  the time-splitting techniques  with Fourier/sine pseudospectral approximation
to simulate the dynamics of rotating dipolar BECs. The numerical methods is explicit, unconditional stable,
spectral accurate in space and second order accurate in time,  and conserves the mass in the discretized level.
The memory cost is $O(MK)$ in 2D and $O(MKL)$ in 3D, and the computational cost per time step
is $O\left(MK\ln(MK)\right)$ in 2D and $O\left(MKL\ln(MKL)\right)$ in 3D.
More specifically, the method is very easy to be implemented via FFT or DST.
We then numerically examine
the conservation of the angular momentum expectation  and study the dynamics of condensate
widths and center of mass for different angular velocities. In addition, the dynamics of vortex lattice
in rotating dipolar BEC are investigated. Numerical studies show that our method is very effective in
simulating the  dynamics of rotating dipolar BECs.

\bigskip

\noindent{\bf Acknowledgement  }
We acknowledge Professor Christof Sparber  for stimulating and helpful
discussions. Part of this work was done when the authors were visiting the
Institute for Mathematical Sciences at the National University of Singapore in 2012.


\end{document}